\documentclass[runningheads]{llncs}
\usepackage{graphicx}

\newif{\ifplots}\plotstrue

\usepackage{todonotes}

\usepackage{enumitem}
\usepackage{mathtools}
\usepackage[utf8]{inputenc}
\usepackage{amssymb,amsmath}
\usepackage{fancyhdr}
\usepackage{stmaryrd}
\usepackage{xspace}
\usepackage{tikz}
\usepackage{multicol}
 \usepackage{pgfplots}
\usetikzlibrary{spy}
\usepackage{pgfplotstable}

\usepackage{algpseudocode}
\usepackage{caption}
\usepackage{algorithmicx}
\usepackage{algorithm}
\usepackage[capitalise]{cleveref}
\usepackage{caption}
\usepackage{subcaption}

\newcommand\ie{i.\,e., }

\newcommand\eg{e.\,g.\xspace}

\newcommand{\Z}{\ensuremath{\mathbb{Z}}}

\newcommand{\R}{\ensuremath{\mathbb{R}}}

\newcommand{\floor}[1]{\left\lfloor\mathinner{#1} \right\rfloor}

\newcommand{\oneset}[1]{\left\{\, \mathinner{#1} \,\right\}}

\begin{document}
\allowdisplaybreaks
\setlength{\abovecaptionskip}{4pt}
\setlength{\belowcaptionskip}{0pt}

\title{On the Average Case of MergeInsertion\thanks{The second author has been supported by the German Research Foundation (DFG) under grant DI 435/7-1.}}
%
%
\author{Florian Stober \and 
	Armin Weiß\orcidID{0000-0002-7645-5867} }
\authorrunning{Florian Stober and 	Armin Weiß}
%
\institute{FMI, Universit{\"a}t Stuttgart, Germany\\
	\email{florian.stober@t-online.de}, 
	\email{armin.weiss@fmi.uni-stuttgart.de}
}
\maketitle              
\begin{abstract}
	MergeInsertion, also known as the Ford-Johnson algorithm, is a sorting algorithm which, up to today, for many input sizes achieves the best known upper bound on the number of comparisons. Indeed, it gets extremely close to the information-theoretic lower bound.
	While the worst-case behavior is well understood, only little is known about the average case.
	This work takes a closer look at the average case behavior.
	In particular, we establish an upper bound of $n \log n - 1.4005n + o(n)$ comparisons.
	We also give an exact description of the probability distribution of the length of the chain a given element is inserted into and use it to approximate the average number of comparisons numerically. Moreover, we compute the exact average number of comparisons for $n$ up to 148.
	Furthermore, we experimentally explore the impact of different decision trees for binary insertion.
	To conclude, we conduct experiments showing that a slightly different insertion order leads to a better average case and we compare the algorithm to the recent combination with (1,2)-Insertionsort by Iwama and Teruyama.


\end{abstract}%
\section{Introduction}\label{sec:intro}
\vspace{-2mm}

Sorting a set of elements is an important operation frequently performed by many computer programs.
Consequently there exist a variety of algorithms for sorting, each of which comes with its own advantages and disadvantages. 

Here we focus on comparison based sorting and study a specific sorting algorithm known as MergeInsertion.
It was discovered by Ford and Johnson in 1959 \cite{FordJohnson}.
Before D. E. Knuth coined the term MergeInsertion in his study of the algorithm in his book ``The Art of Computer Programming,  Volume 3: Sorting and Searching'' \cite{Knuth}, it was known only as Ford-Johnson Algorithm, named after its creators.
The one outstanding property of MergeInsertion is that the number of comparisons it requires is close to the information-theoretic lower bound of $\log(n!) \approx n \log n - 1.4427n$ (for sorting $n$ elements).
This sets it apart from many other sorting algorithms.
MergeInsertion can be described in three steps:
first pairs of elements are compared;
in the second step the larger elements are sorted recursively;
as a last step the elements belonging to the smaller half are inserted into the already sorted larger half using binary insertion.

In the worst case the number of comparisons of MergeInsertion is quite well understood \cite{Knuth}~-- it is $n \log n + b(n)\cdot n + o(n)$ where $b(n)$ oscillates between $-1.415$ and $-1.3289$. Moreover, for many $n$ MergeInsertion is proved to be the optimal algorithm in the worst case (in particular, for $n \leq 15$ \cite{Peczarski04,Peczarski07}). However, there are also $n$ where it is not optimal \cite{Manacher,bui1985significant}. One reason for this is the oscillating linear term in the number of comparisons, which allowed Manacher \cite{Manacher} to show that for certain $n$ it is more efficient to split the input into two parts, sort both parts with MergeInsertion, and then merge the two parts into one array.

Regarding the average case not much is known: in \cite{Knuth} Knuth calculated the number of comparisons required on average for $n \in \lbrace 1, \dots, 8 \rbrace$;
an upper bound of $n \log n -1.3999n +o(n)$ has been established in \cite{EdelkampW14}. Most recently, Iwama and Teruyama \cite{12ins} showed that in the average case MergeInsertion can be improved by combining it with their (1,2)-Insertion algorithm resulting in an upper bound of $n \log n -1.4106n +O(\log n)$. This reduces the gap to the lower bound by around 25\%. It is a fundamental open problem how close one can get to the information-theoretic lower bound of $n \log n - 1.4427n$ (see \eg \cite{12ins,Reinhardt92}).

The goal of this work is to study the number of comparisons required in the average case.
In particular, we analyze the insertion step of MergeInsertion in greater detail.
In general, MergeInsertion achieves its good performance by inserting elements in a specific order that in the worst case causes each element to be inserted into a sorted list of $2^k-1$ elements (thus, using exactly $k$ comparisons). When looking at the average case elements are often inserted into less than $2^k-1$ elements which is slightly cheaper.
By calculating those small savings we seek to achieve our goal of a better upper bound on the average case.
Our results can be summarized as follows:
\begin{itemize}[noitemsep, topsep=0pt]
	\item We derive an exact formula for the probability distribution into how many elements a given element is inserted (\cref{th:Yi}). This is the crucial first step in order to obtain better bounds for the average case of MergeInsertion.
	\item We experimentally examine different decision trees for binary insertion. We obtain the best result when assigning shorter decision paths to positions located further to the left.
	\item We use \cref{th:Yi} in order to compute quite precise numerical estimates for the average number of comparisons for $n$ up to roughly 15000.
	\item We compute the exact average number of comparisons for $n$ up to 148~-- thus, going much further than \cite{Knuth}.
	\item We improve the bound of \cite{EdelkampW14} to $n \log n - 1.4005n + o(n)$ (\cref{thm:upper_bound}). This partially answers a conjecture from \cite{Reinhardt92} which asks for an in-place algorithm with $n\log n + 1.4n$ comparisons on average and $n\log n - 1.3 n$ comparisons in the worst case. Although MergeInsertion is not in-place, the the techniques from \cite{EdelkampW14} or \cite{Reinhardt92} can be used to make it so.
	
	\item We evaluate a slightly different insertion order decreasing the gap between the lower bound and the average number of comparisons of MergeInsertion by roughly 30\% for $n\approx 2^k/3$.  
	\item We compare MergeInsertion to the recent combination by Iwama and Teruyama \cite{12ins} showing that, in fact, their combined algorithm is still better than the analysis and with the different insertion order can be further improved.
\end{itemize}
Most proofs as well as additional explanations and experimental results can be found in the appendix.
The code used in this work and the generated data is available on \cite{code}.

\section{Preliminaries}\label{sec:prelims}
Throughout, we assume that the input consists of $n$ distinct elements.
The average case complexity is the mean number of comparisons over all input permutations of $n$ elements.

\subsubsection*{Description of MergeInsertion}
The MergeInsertion algorithm consists of three phases: pairwise comparison, recursion, and insertion. 
Accompanying the explanations we give an example where $n = 21$. We call such a set of relations between individual elements a configuration.%
\begin{enumerate}

	\item \textbf{Pairwise comparison}.
	The elements are grouped into $\floor{\frac{n}{2}}$ pairs.
	Each pair is sorted using one comparison. 
	After that, the elements are called $a_1$ to $a_{\left\lfloor \frac{n}{2} \right\rfloor}$ and $b_1$ to $b_{\left\lceil \frac{n}{2} \right\rceil }$ with $ a_i > b_i$ for all $1\le i\le \left\lfloor \frac{n}{2} \right\rfloor$.
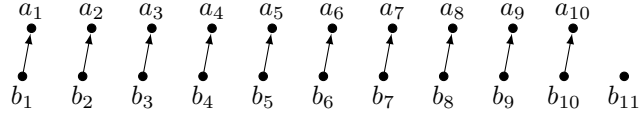
		\begin{center}
			\begin{tikzpicture}[>=latex,scale=.8]
			\foreach \x in {1, 2, ..., 10}
			\filldraw (\x - 0.15,0) circle (2pt) node [below] {$b_{\x}$}
			(\x, 0.8) circle (2pt) node [above] {$a_{\x}$};
			\foreach \x in {1, 2, ..., 10}
			\draw[->,shorten >=2pt] (\x - 0.15,0) -- (\x, 0.8);
			\filldraw (11 - 0.15,0) circle (2pt) node [below] {$b_{11}$};
			\end{tikzpicture}
		\end{center}

	\item \textbf{Recursion}.
	The $\left\lfloor \frac{n}{2} \right\rfloor$ larger elements, \ie $a_1$ to $a_{\left\lfloor \frac{n}{2} \right\rfloor}$ are sorted recursively.
	Then all elements (the $\left\lfloor \frac{n}{2} \right\rfloor$ larger ones as well as the corresponding smaller ones) are renamed accordingly such that $a_i < a_{i+1}$ and $a_i > b_i$ still holds.
		
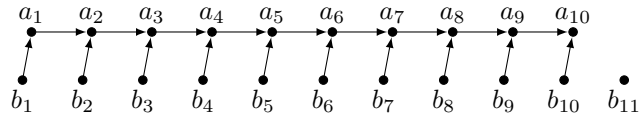
\begin{center}
			\begin{tikzpicture}[>=latex,scale=.8]
			\foreach \x in {1, 2, ..., 10}
			\filldraw(\x - 0.15,0) circle (2pt) node [below] {$b_{\x}$}
			(\x, 0.8) circle (2pt) node [above] {$a_{\x}$};
			\foreach \x in {1, 2, ..., 10}
			\draw[-latex,shorten >=2pt] (\x - 0.15,0) -- (\x, 0.8);
			\filldraw (11 - 0.15,0) circle (2pt) node [below] {$b_{11}$};
			\foreach \x in {1, 2, ..., 9}
			\draw[-latex,shorten >=2pt] (\x, 0.8) -- (\x + 1, 0.8);
			\end{tikzpicture}
		\end{center}

	\item \textbf{Insertion}. The $\left\lceil \frac{n}{2} \right\rceil$ small elements, \ie the $b_i$, are inserted into the main chain using binary insertion.
	The term ``main chain'' describes the set of elements containing $a_1, \dots,a_{t_k}$ as well as the $b_i$ that have already been inserted.

	The elements are inserted in batches starting with $b_3, b_2$.
	In the $k$-th batch the elements $b_{t_k}, b_{t_k - 1}, \dots, b_{t_{k - 1} + 1}$ where $t_k = \frac{2^{k+1} + (-1)^k}{3}$ are inserted in that order.
	Elements $b_j$ where $j > \left \lceil \frac{n}{2} \right \rceil$ (which do not exist) are skipped. 
  Note that technically $b_1$ is the first batch; but inserting $b_1$ does not need any comparison.

	Because of the insertion order, every element $b_i$ which is part of the $k$-th batch is inserted into at most $2^{k} - 1$ elements; thus, it can be inserted by binary insertion using at most $k$ comparisons.
			
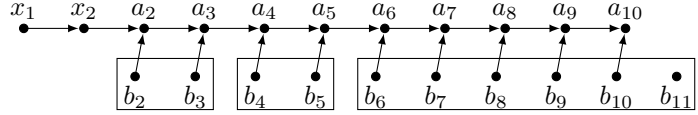
\begin{center}
				\begin{tikzpicture}[>=latex,scale=.8]
				\foreach \x in {2, 3, ..., 10}
				\filldraw(\x - 0.15,0) circle (2pt) node [below] {$b_{\x}$}
				(\x, 0.8) circle (2pt) node [above] {$a_{\x}$};
				\foreach \x in {2, 3, ..., 10}
				\draw(\x - 0.15,0) [-latex,shorten >=2pt]-- (\x, 0.8);
				\filldraw (11 - 0.15,0) circle (2pt) node [below] {$b_{11}$};
				\filldraw (0, 0.8) circle (2pt) node [above] {$x_1$};
				\filldraw (1, 0.8) circle (2pt) node [above] {$x_2$};
				\foreach \x in {0, 1, ..., 9}
				\draw[-latex,shorten >=2pt] (\x, 0.8) -- (\x + 1, 0.8);
				\draw (1.55, -0.55) rectangle (3.15, 0.3) ;
				\draw (3.55, -0.55) rectangle (5.15, 0.3) ;
				\draw (5.55, -0.55) rectangle (11.15, 0.3) ;
				\end{tikzpicture}
			\end{center}
\end{enumerate}

Regarding the average number of comparisons $F(n)$ we make the following observations:
the first step always requires $\left\lfloor \frac{n}{2} \right\rfloor$ comparisons.
The recursion step does not do any comparisons by itself but depends on the other steps.
The average number of comparisons $G(n)$ required in the insertion step is not obvious.
It will be studied closer in following chapters.
Following \cite{Knuth}, we obtain the recurrence  (which is the same as for the worst-case number of comparisons)
\begin{equation}
F(n) = \left\lfloor \frac{n}{2} \right\rfloor + F\left(\left\lfloor \frac{n}{2} \right\rfloor\right) + G\left(\left\lceil \frac{n}{2} \right\rceil\right).
\label{eq:Fn}
\end{equation}

\section{Average Case Analysis of the Insertion Step}\label{sec:avg_analysis}

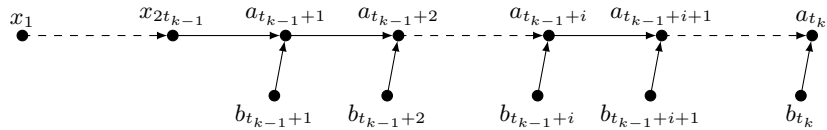
\begin{figure}[h]
	\centering
	\vspace{-5mm}
	\begin{tikzpicture}
	\filldraw (0, 0.8) circle (2pt) node [above] {$x_1$}
	(2, 0.8) circle (2pt) node [above] {$x_{2t_{k-1}}$}
	(3.5, 0.8) circle (2pt) node [above] {$a_{t_{k-1}+1}$}
	(5, 0.8) circle (2pt) node [above] {$a_{t_{k-1}+2}$}
	(7, 0.8) circle (2pt) node [above] {$a_{t_{k-1}+i}$}
	(8.5, 0.8) circle (2pt) node [above] {$a_{t_{k-1}+i+1}$}
	(10.5, 0.8) circle (2pt) node [above] {$a_{t_k}$};

	\draw[-latex,shorten >=2pt](2, 0.8) -- (3.5, 0.8);
	\draw[-latex,shorten >=2pt](3.5, 0.8) -- (5, 0.8);
	\draw[-latex,shorten >=2pt](7, 0.8) -- (8.5, 0.8);
	\draw[-latex,shorten >=2pt,dashed] (0, 0.8) -- (2, 0.8);
	\draw[-latex,shorten >=2pt,dashed] (5, 0.8) -- (7, 0.8);
	\draw[-latex,shorten >=2pt,dashed] (8.5, 0.8) -- (10.5, 0.8);

	\filldraw (3.5 - 0.15, 0) circle (2pt) node [below] {$b_{t_{k-1}+1}$}
	(5 - 0.15, 0) circle (2pt) node [below] {$b_{t_{k-1}+2}$}
	(7 - 0.15, 0) circle (2pt) node [below] {$b_{t_{k-1}+i}$}
	(8.5 - 0.15, 0) circle (2pt) node [below] {$b_{t_{k-1}+i+1}$}
	(10.5 - 0.15, 0) circle (2pt) node [below] {$b_{t_k}$};

	\draw[-latex,shorten >=2pt](3.5 - 0.15,0) -- (3.5, 0.8);
	\draw[-latex,shorten >=2pt](5 - 0.15,0) -- (5, 0.8);
	\draw[-latex,shorten >=2pt](7 - 0.15,0) -- (7, 0.8);
	\draw[-latex,shorten >=2pt](8.5 - 0.15,0) -- (8.5, 0.8);
	\draw[-latex,shorten >=2pt](10.5 - 0.15,0) -- (10.5, 0.8);
	\end{tikzpicture}
	\caption{Configuration where a single batch of elements remains to be inserted}\label{fig:ins_bi}
\end{figure}

In this section we have a look at different probabilities when inserting one batch of elements, \ie the elements $b_{t_k}$ to $b_{t_{k-1}+1}$. 
 We assume that all elements of previous batches, \ie $b_1$ to $b_{t_{k-1}}$, have already been inserted and together with the corresponding $a_i$ they constitute the main chain and have been renamed to $x_1$ to $x_{2t_{k-1}}$ such that $x_i < x_{i+1}$. The situation is shown in \cref{fig:ins_bi}.

We will look at the element $b_{t_k+i}$ and want to answer the following questions:
what is the probability of it being inserted between $x_j$ and $x_{j+1}$?
And what is the probability of it being inserted into a specific number of elements?

\vspace{-3mm}

We can ignore batches that are inserted after the batch we are looking at since those do not affect the probabilities we want to obtain.

First we define a probability space for the process of inserting one batch of elements: let
$\Omega_k$ be the set of all possible outcomes (\ie linear extensions) when sorting the partially ordered elements shown in \cref{fig:ins_bi} by inserting $b_{t_k}$ to $b_{t_{k-1}+1}$.
Each $\omega \in \Omega_k$ can be viewed as a function that maps an element $e$ to its final position, \ie $\omega(e) \in \left \lbrace 1, 2, \dots, 2t_k \right \rbrace$.
While the algorithm mandates a specific order for inserting the elements $b_{t_{k-1}+1}$ to $b_{t_k}$ during the insertion step, using a different order does not change the outcome, \ie the elements are still sorted correctly.
For this reason we can assume a different insertion in order to simplify calculating the likelihood of relations between individual elements.

Let us look at where an element will end up after it has been inserted.
Not all positions are equally likely.
For this purpose we define the random variable $X_i$ as follows. To simplify notation we define $x_{t_{k-1} + j} := a_{j}$ for  $ t_{k-1} < j\leq  t_k $ (hence, the main chain consists of $x_1, \dots, x_{2^k}$). 
\begin{equation*}
X_i : \omega \mapsto \left\{\begin{array}{llr}
0 & \text{if } \omega(b_{t_{k-1}+i}) < \omega(x_1) & \\
j & \text{if } \omega(x_j)  < \omega(b_{t_{k-1}+i}) < \omega(x_{j+1}) & \text{for } j \in \{1, \dots, 2^k - 2\} \\
2^k - 1 & \text{if } \omega(x_{2^k - 1}) < \omega(b_{t_{k-1}+i})\text{.} &
\end{array}\right.
\end{equation*}
We are interested in the probabilities $P(X_i = j)$. %
These values follow a simple pattern (for $k = 4$ these are given  
in \cref{tab:insert} in the appendix).

\begin{theorem}
	\label{th:Xi}
	The probability of $b_{t_{k-1}+i}$ being inserted between $x_j$ and $x_{j+1}$ is given by
	\vspace{-4mm}\begin{equation*}
	P(X_i = j) =  \left\{\begin{array}{ll}
	2^{2i-2} \left(\frac{\left(t_{k-1}+i-1\right)!}{\left(t_{k-1}\right)!}\right)^2 \frac{\left(2t_{k-1}\right)!}{\left(2t_{k-1}+2i-1\right)!} & \text{if } 0 \leq j \leq 2t_{k-1}  \\
	2^{4t_{k-1}-2j+2i-2} \!\left(\!\frac{\left(t_{k-1}+i-1\right)!}{\left(j-t_{k-1}\right)!}\!\right)^{\!\!2\vphantom{2^{2^2}}} \!\!\!\frac{\left(2j-2t_{k-1}\right)!}{\left(2t_{k-1}+2i-1\right)!} & \text{if } 2t_{k-1} < j < 2t_{k-1} \!\!+\! i \\
	0 & \text{otherwise}
	\end{array}\right.
	\end{equation*}
\end{theorem}

Next, our aim is to compute the probability that $b_i$ is inserted into a particular number of elements. This is of particular interest because the difference between average and worst case comes from the fact that sometimes we insert into less than $2^k-1$ elements.
For that purpose we define the random variable $Y_i$.
	\begin{equation*}
Y_i: \omega \mapsto \left\vert \left\{ v \in \oneset{x_1, \dots, x_{2^k} } \cup \{ b_{t_{k-1}+i+ 1}, \dots, b_{t_{k}}\}  \mid \omega(v) < \omega(a_{t_{k-1}+i}) \right\} \right\vert
\end{equation*}
The elements in the main chain when inserting $b_{t_k+i}$ are $x_1$ to $x_{2t_{k-1} + i-1}$
 and those elements out of ${b_{t_{k-1}+i+1}}, \dots, b_{t_k}$ which have been inserted before $a_{t_{k-1}+i}$ (which is $x_{2t_{k-1}+i}$).
For computing the number of these, we introduce random variables $\tilde{Y}_{i,q}$ counting the elements in $\{b_{t_{k-1}+i+1}, \dots, b_{t_{k-1}+i+q}\}$ that are inserted before $a_{t_{k-1}+i}$:
\begin{equation*}
\tilde{Y}_{i,q}: \omega \mapsto \left\vert \left\{  v \in \{b_{t_{k-1}+i+1}, \dots, b_{t_{k-1}+i+q}\}\mid \omega(v) < \omega(a_{t_{k-1}+i})  \right\} \right\vert.
\end{equation*}
By setting $q=t_k-t_{k-1}-i$, we obtain 
$
Y_i = \tilde{Y}_{i,t_k-t_{k-1}-i} + 2t_{k-1} + i - 1.
$
For an illustration see \Cref{fig:proof_px_cond} in the appendix. 
Clearly we have $P\bigl(\tilde{Y}_{i,0} = j\bigr) = 1 $ if $j=0$ and $P\bigl(\tilde{Y}_{i,0} = j\bigr) = 0 $ otherwise.
 For $q>0$ there are two possibilities:
\begin{enumerate}
	\item $\tilde{Y}_{i,q-1} = j-1$ and $X_{i+q} < 2t_{k-1}+i$: out of $\{b_{t_{k-1}+i+1}, \dots, b_{t_{k-1}+i+q-1}\}$ there have been $j-1$ elements inserted before $a_{t_{k-1}+i}$ and $b_{t_{k-1}+i+q}$ is inserted before $a_{t_{k-1}+i}$.
	\item $\tilde{Y}_{i,q-1} = j$ and $X_{i+q} \ge 2t_{k-1}+i$:  out of $\{b_{t_{k-1}+i+1}, \dots, b_{t_{k-1}+i+q-1}\}$ there have been $j$ elements inserted before $a_{t_{k-1}+i}$ and $b_{t_{k-1}+i+q}$ is inserted after $a_{t_{k-1}+i}$.
\end{enumerate}%

From these we obtain the following recurrence:
\begin{align*}
P(\tilde{Y}_{i,q} = j) 
=\quad&P(X_{i+q} < 2t_{k-1}+i \:\vert\: \tilde{Y}_{i,q-1}=j-1)\cdot P(\tilde{Y}_{i,q-1}=j-1)\\
+&P(X_{i+q} \ge 2t_{k-1}+i \:\vert\: \tilde{Y}_{i,q-1}=j)\cdot P(\tilde{Y}_{i,q-1}=j)
\end{align*}

The probability $P(X_{i+q} < 2t_{k-1}+i \:\vert\: \tilde{Y}_{i,q-1}=j-1)$ can be obtained by looking at \cref{fig:ins_bi} and counting elements.
When $b_{t_{k-1}+i+q}$ is inserted, the elements on the main chain which are smaller than $a_{t_{k-1}+i}$ are $x_1$ to $x_{2t_{k-1}}$, $a_{t_{k-1}+1}$ to $a_{t_{k-1}+i-1}$ and $j-1$ elements out of $\{b_{t_{k-1}+i+1}, \dots, b_{t_{k-1}+i+q-1}\}$ which is a total of $2t_{k-1}+2i+j-2$ elements.
Combined with the fact that the main chain consists of $2t_{k-1} + 2i + 2q - 2$ elements smaller than $a_{t_{k-1}+i+q}$ we obtain the probability $\frac{2t_{k-1}+2i+j-1}{2t_{k-1} + 2i + 2q - 1}$.
We can calculate $P(X_{i+q} \ge 2t_{k-1}+i \:\vert\: \tilde{Y}_{i,q-1}=j)$ similarly leading to
\begin{equation*}
\scalebox{0.9}{$
P(\tilde{Y}_{i,q} = j)=\frac{2t_{k-1}+2i+j-1}{2t_{k-1}+2i+2q-1}\cdot P(\tilde{Y}_{i,q-1}=j-1)+\frac{2q-j-1}{2t_{k-1}+2i+2q-1}\cdot P(\tilde{Y}_{i,q-1}=j)
$}.
\end{equation*}

By solving the recurrence, we obtain a closed form for $P(\tilde{Y}_{i,q} = j)$ and, thus, for $P(Y_i = j)$.
 The complete proof is given in \Cref{proof:Yi}.

\begin{theorem}
	\label{th:Yi}
	For $1 \le i \le t_k-t_{k-1}$ and $2t_{k-1}+i-1 \le j \le 2^k-1$ the probability $P(Y_i = j)$, that $b_{t_{k-1}+i}$ is inserted into $j$ elements is given by
	\begin{equation*}
	\scalebox{0.93}{$\displaystyle
		P(Y_i = j) = 2^{j-2t_{k-1}-i+1}\frac{(2t_k-i-j-1)!}{(j-2t_{k-1}-i+1)!(2^k-j-1)!}\frac{(i+j)!}{(2t_k-1)!}\frac{(t_k-1)!}{(t_{k-1}+i-1)}.
		$}
	\end{equation*}
\end{theorem}

\Cref{fig:distr_pYi} shows the probability distribution for $Y_1$, $Y_{21}$ and $Y_{42}$ where $k=7$.
$Y_{42}$ corresponds to the insertion of $b_{t_k}$ (the first element of the batch).
$Y_1$ corresponds to the insertion of $b_{t_{k-1}+1}$ (the last element of the batch).
In addition to those three probability distributions \cref{fig:distr_eYi} shows the mean of all $Y_i$ for $k=7$.

\ifplots

\begin{figure}[t]
	\begin{minipage}[b]{0.5\textwidth}
		\begin{tikzpicture}
		\begin{axis}[xlabel=j,
		ylabel=probability,
		legend pos = north west,
		width=\textwidth,
		height=\axisdefaultheight*0.65]
		\addplot[green,mark=*,mark repeat={1},mark size=0.75] table [x=j, y=Y1,  col sep=tab] {distr_pYi_k7.csv};
		\addlegendentry{$P(Y_1=j)$}
		\addplot[blue,mark=*,mark repeat={1},mark size=0.75] table [x=j, y=Y21, col sep=tab] {distr_pYi_k7.csv};
		\addlegendentry{$P(Y_{21}=j)$}
		\addplot[red,mark=*,mark repeat={1},mark size=0.75] table [x=j, y=Y42, col sep=tab] {distr_pYi_k7.csv};
		\addlegendentry{$P(Y_{42}=j)$}
		\end{axis}
		\end{tikzpicture}
		\caption{Probability distribution of $Y_i$.}
		\label{fig:distr_pYi}
	\end{minipage}%
	\begin{minipage}[b]{0.5\textwidth}
		\begin{tikzpicture}
		\begin{axis}[xlabel=$i\vphantom{j}$,
		ylabel=$E(Y_i)$,
		width=\textwidth,
		height=\axisdefaultheight*0.65]
		\addplot[blue,mark=*,mark repeat={1},mark size=0.75] table [x=i, y=EYi, col sep=tab] {distr_eYi_k7.csv};
		\end{axis}
		\end{tikzpicture}
		\caption{Mean of $Y_i$ for different $i$. $k=7$.}
		\label{fig:distr_eYi}
	\end{minipage}
\end{figure}
\fi

\subsubsection*{Binary Insertion and different decision trees}
\label{chap:BinIns}

\begin{figure}[t]
	\begin{subfigure}[c]{0.25\textwidth}
		\begin{tikzpicture}[>=latex, scale=.52]
		\foreach \x in {1, 2, ..., 5}
		\filldraw(\x,0) circle (2pt) node [below] {$\x$};
		\draw[dashed] (1, 1) -- (1, 0);
		\draw[dashed] (2, 0.5) -- (2, 0);
		\draw[dashed] (3, 1.5) -- (3, 0);
		\draw[dashed] (4, 1) -- (4, 0);
		\draw[dashed] (5, 0.5) -- (5, 0);
		\draw[->] (3, 1.5) -- (1, 1);
		\draw[->] (3, 1.5) -- (4, 1);
		\draw[->] (1, 1) -- (2, 0.5);
		\draw[->] (4, 1) -- (5, 0.5);
		\draw[->, rounded corners] (1, 1) -| (0.5, 0);
		\draw[->, rounded corners] (2, 0.5) -| (1.5, 0);
		\draw[->, rounded corners] (2, 0.5) -| (2.5, 0);
		\draw[->, rounded corners] (4, 1) -| (3.5, 0);
		\draw[->, rounded corners] (5, 0.5) -| (4.5, 0);
		\draw[->, rounded corners] (5, 0.5) -| (5.5, 0);
		\end{tikzpicture}
		\caption{\texttt{center-left}}
		\label{fig:BinIns_CL}
	\end{subfigure}%
	\begin{subfigure}[c]{0.25\textwidth}
		\begin{tikzpicture}[>=latex, scale=.52]
		\foreach \x in {1, 2, ..., 5}
		\filldraw(\x,0) circle (2pt) node [below] {$\x$};
		\draw[dashed] (1, 0.5) -- (1, 0);
		\draw[dashed] (2, 1) -- (2, 0);
		\draw[dashed] (3, 1.5) -- (3, 0);
		\draw[dashed] (4, 0.5) -- (4, 0);
		\draw[dashed] (5, 1) -- (5, 0);
		\draw[->] (3, 1.5) -- (2, 1);
		\draw[->] (3, 1.5) -- (5, 1);
		\draw[->] (2, 1) -- (1, 0.5);
		\draw[->] (5, 1) -- (4, 0.5);
		\draw[->, rounded corners] (1, 0.5) -| (0.5, 0);
		\draw[->, rounded corners] (1, 0.5) -| (1.5, 0);
		\draw[->, rounded corners] (2, 1) -| (2.5, 0);
		\draw[->, rounded corners] (4, 0.5) -| (3.5, 0);
		\draw[->, rounded corners] (4, 0.5) -| (4.5, 0);
		\draw[->, rounded corners] (5, 1) -| (5.5, 0);
		\end{tikzpicture}
		\caption{\texttt{center-right}}
		\label{fig:BinIns_CR}
	\end{subfigure}%
	\begin{subfigure}[c]{0.25\textwidth}
		\begin{tikzpicture}[>=latex, scale=.52]
		\foreach \x in {1, 2, ..., 5}
		\filldraw(\x,0) circle (2pt) node [below] {$\x$};
		\draw[dashed] (1, 1) -- (1, 0);
		\draw[dashed] (2, 1.5) -- (2, 0);
		\draw[dashed] (3, 0.5) -- (3, 0);
		\draw[dashed] (4, 1) -- (4, 0);
		\draw[dashed] (5, 0.5) -- (5, 0);
		\draw[->] (2, 1.5) -- (1, 1);
		\draw[->] (2, 1.5) -- (4, 1);
		\draw[->] (4, 1) -- (3, 0.5);
		\draw[->] (4, 1) -- (5, 0.5);
		\draw[->, rounded corners] (1, 1) -| (0.5, 0);
		\draw[->, rounded corners] (1, 1) -| (1.5, 0);
		\draw[->, rounded corners] (3, 0.5) -| (2.5, 0);
		\draw[->, rounded corners] (3, 0.5) -| (3.5, 0);
		\draw[->, rounded corners] (5, 0.5) -| (4.5, 0);
		\draw[->, rounded corners] (5, 0.5) -| (5.5, 0);
		\end{tikzpicture}
		\caption{\texttt{left}}
		\label{fig:BinIns_L}
	\end{subfigure}%
	\begin{subfigure}[c]{0.25\textwidth}
		\begin{tikzpicture}[>=latex, scale=.52]
		\foreach \x in {1, 2, ..., 5}
		\filldraw(\x,0) circle (2pt) node [below] {$\x$};
		\draw[dashed] (1, 0.5) -- (1, 0);
		\draw[dashed] (2, 1) -- (2, 0);
		\draw[dashed] (3, 0.5) -- (3, 0);
		\draw[dashed] (4, 1.5) -- (4, 0);
		\draw[dashed] (5, 1) -- (5, 0);
		\draw[->] (4, 1.5) -- (2, 1);
		\draw[->] (4, 1.5) -- (5, 1);
		\draw[->] (2, 1) -- (1, 0.5);
		\draw[->] (2, 1) -- (3, 0.5);
		\draw[->, rounded corners] (1, 0.5) -| (0.5, 0);
		\draw[->, rounded corners] (1, 0.5) -| (1.5, 0);
		\draw[->, rounded corners] (3, 0.5) -| (2.5, 0);
		\draw[->, rounded corners] (3, 0.5) -| (3.5, 0);
		\draw[->, rounded corners] (5, 1) -| (4.5, 0);
		\draw[->, rounded corners] (5, 1) -| (5.5, 0);
		\end{tikzpicture}
		\caption{\texttt{right}}
		\label{fig:BinIns_R}
	\end{subfigure}%
	\caption{Different strategies for binary insertion.\vspace{-1mm}}
	\label{fig:BinIns}
\end{figure}
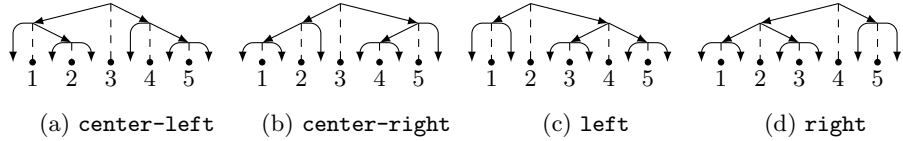

The Binary Insertion step is an important part of MergeInsertion.
In the average case many elements are inserted in less than $2^k-1$ (which is the worst case).
This leads to ambiguous decision trees where at some positions inserting an element requires only $k-1$ instead of $k$ comparisons. Since not all positions are equally likely (positions on the left have a slightly higher probability), this results in different average insertion costs.
We compare four different strategies all satisfying that the corresponding decision trees have their leaves distributed across at most two layers.
For an example with five elements see \Cref{fig:BinIns}.

First there are the \texttt{center-left} and \texttt{center-right} strategies (the standard options for binary insertion):
they compare the element to be inserted with the middle element, rounding down(up) in case of an odd number.
The \texttt{left} strategy chooses the element to compare with in a way such that the positions where only $k-1$ comparisons are required are at the very left.
The \texttt{right} strategy is similar, here the positions where one can insert with just $k-1$ comparisons are at the right.
To summarize, the element to compare with is%
\begin{center}
	\small
	\begin{tabular}{ll}
		$\left\lfloor\frac{n+1}{2}\right\rfloor$ & strategy \texttt{center-left}\\
		$\left\lceil\frac{n+1}{2}\right\rceil\vphantom{2^{k^k}}$ & strategy \texttt{center-right}\\
		$\max\lbrace n-2^k+1, 2^{k-1}\rbrace\quad$ & strategy \texttt{left}\\
		$	\min\lbrace 2^k, n - 2^{k-1}+1\rbrace$ & strategy \texttt{right}
	\end{tabular}
\end{center}%
where $k = \lfloor \log n \rfloor$. 
Notice that the \texttt{left} strategy is also used in \cite{12ins}, where it is called \emph{right-hand-binary-search}. 
\Cref{fig:bin_ins_results} shows experimental results comparing the different strategies for binary insertion regarding their effect on the average-case of MergeInsertion.
As we can see the \texttt{left} strategy performs the best, closely followed by \texttt{center-left} and \texttt{center-right}.
\texttt{right} performs the worst.
The \texttt{left} strategy performing best is no surprise since
the probability that an element is inserted into one of the left positions is higher that it being inserted to the right.
Therefore, in all further experiments we use the \texttt{left} strategy.

\usetikzlibrary{spy}
\begin{figure}[t]
		\begin{tikzpicture}[spy using outlines={circle, magnification=6, connect spies}]
		\begin{axis}[xlabel=number of elements $n$,
		ylabel=$\frac{\text{number of comparisons} - n\log n}{n}$,
		legend pos = north west,
		xmode=log,
		log basis x=2,
		ytick={-1.435, -1.430, -1.425, -1.420, -1.415, -1.410},
		yticklabels={-1.435, -1.430, -1.425, -1.420, -1.415, -1.410},
		width=\textwidth,
		height=0.8*\axisdefaultheight]
		\addplot[green,mark=*,mark repeat={2},mark size=0.5] table [x=num_elements, y=center-left, col sep=tab] {binary_insertion.csv};
		\addlegendentry{\texttt{center-left}}
		\addplot[blue,mark=*,mark repeat={2},mark size=0.5] table [x=num_elements, y=center-right, col sep=tab] {binary_insertion.csv};
		\addlegendentry{\texttt{center-right}}
		\addplot[red,mark=*,mark repeat={2},mark size=0.5] table [x=num_elements, y=left, col sep=tab] {binary_insertion.csv};
		\addlegendentry{\texttt{left}}
		\addplot[brown,mark=*,mark repeat={2},mark size=0.5] table [x=num_elements, y=right, col sep=tab] {binary_insertion.csv};
		\addlegendentry{\texttt{right}}
		\coordinate (spypoint) at (axis cs:87500,-1.4325);
		\coordinate (magnifyglass) at (axis cs:165000,-1.426);
		\coordinate (spypoint2) at (axis cs:524000,-1.415);
		\coordinate (magnifyglass2) at (axis cs:1200000,-1.425);
		\end{axis}
		\spy [black, size=2.5cm] on (spypoint)
		in node[fill=white] at (magnifyglass);
		\spy [black, size=2.5cm] on (spypoint2)
		in node[fill=white] at (magnifyglass2);
		\end{tikzpicture}
	\caption{Experimental results on the effect of different strategies for binary insertion on the number of comparisons.\vspace{-3mm}}
	\label{fig:bin_ins_results}
\end{figure}
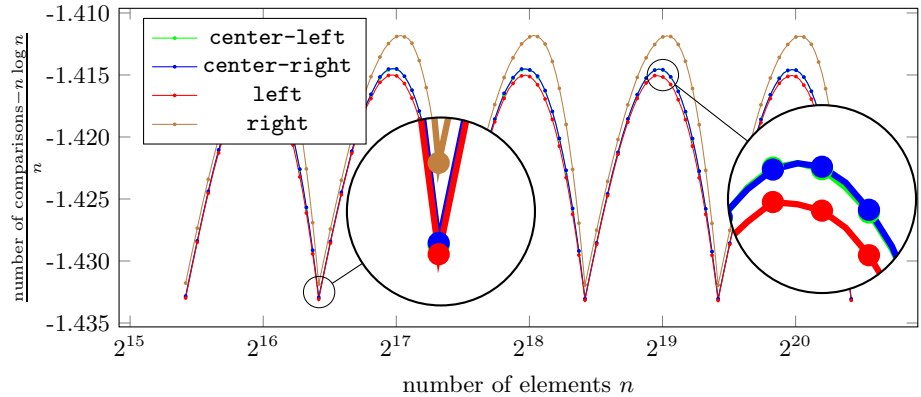

\vspace{-2mm}
\section{Improved Upper Bounds for MergeInsertion}\label{sec:avg_analysis}

\vspace{-2mm}
\subsubsection*{Numeric upper bound}
\label{chap:numeric_upper_bound}

The goal of this section is to combine the probability given by \Cref{th:Yi} that an element $b_{t_{k-1}+i}$ is inserted into $j$ elements with an upper bound for the number of comparisons required for binary insertion.

By \cite{EdelkampWW18}, the number of comparisons required for binary insertion when inserting into $m-1$ elements is $T_{\text{InsAvg}}(m) = \lceil \log m \rceil + 1 - \frac{2^{\lceil \log m \rceil}}{m}$.
While only being exact in case of a uniform distribution, this formula acts as an upper bound in our case, where the probability is monotonically decreasing with the index.

This leads to an upper bound for the cost of inserting $b_{t_{k-1}+i}$ of $T_{\text{Ins}}(i, k) = \sum_j P(Y_i = j)\cdot T_{\text{InsAvg}}(j+1)$.
From there we calculated an upper bound for MergeInsertion.
\Cref{fig:approx} compares those results with experimental data on the number of comparisons required by MergeInsertion.
We observe that the difference is rather small.

\ifplots
\begin{figure}[t]
		\begin{tikzpicture}[spy using outlines={circle, magnification=7, connect spies}]
		\begin{axis}[xlabel=$n$,
		ylabel=$\frac{\text{number of comparisons} - n\log n}{n}$,
		cycle list name=color list,
		legend pos = north east,
		xmode=log,
		log basis x=2,
		width=\textwidth,
		height=0.7*\axisdefaultheight]
		\addplot table [x=num_elements, y=MI, col sep=tab] {mi_small.csv};
		\addlegendentry{Experimental Data}
		\addplot table [x=num_elements, y=approx, col sep=tab] {approx2.csv};
		\addlegendentry{Numeric Upper Bound}
		\coordinate (magnifyglass) at (axis cs:850,-1.39);
		\coordinate (spypoint) at (axis cs:2048,-1.413);
		\end{axis}
		\spy [black, size=2cm] on (spypoint)
		in node[fill=white] at (magnifyglass);
		\end{tikzpicture}
	\caption{Comparing our upper bound with experimental data on the number of comparisons required by MergeInsertion.}
	\label{fig:approx}
\end{figure}
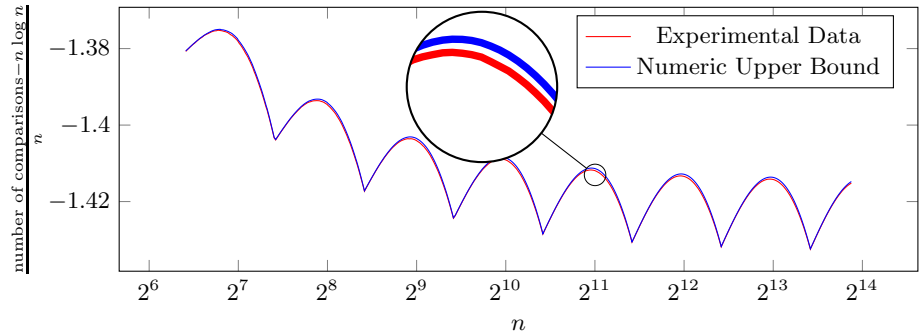
\fi

\vspace{-3mm}
\subsubsection*{Computing the Exact Number of Comparisons}
\label{chap:exact}

In this section we explore how to numerically calculate the exact number of comparisons required in the average case.
The most straightforward way of doing this is to compute the external path length of the decision tree (sum of lengths of all paths from the root to leaves) and dividing by the number of leaves ($n!$ when sorting $n$ elements), which unfortunately is only feasible for very small $n$.
Instead we use \Cref{eq:Fn}, which describes the number of comparisons.
The only unknown in that formula is $G(n)$ the number of comparisons required in the insertion step of the algorithm.
Since the insertion step of MergeInsertion works by inserting elements in batches, we write 
$
G(n) = \left(\sum_{1<k\le k_n} \texttt{Cost}(t_{k-1}, t_{k})\right) + \texttt{Cost}(t_{k_n}, n)$  for $t_{k_n} \le n < t_{k_n+1}
$.
Here $\texttt{Cost}(s, e)$ is the cost of inserting one batch of elements starting from $b_{s+1}$ up to $b_e$.
The idea for computing $\texttt{Cost}(s, e)$ is to calculate the external path length of the decision tree corresponding to the insertion of that batch of elements and then dividing by the number of leaves. As this is still not feasible, we apply some optimizations which we describe in detail in \Cref{chap:appendix_exact}.

For $n \in \lbrace 1, \dots, 15 \rbrace$ the computed values are shown in \cref{tab:exactf}, for larger $n$ \cref{fig:exactf} shows the values we computed.
The complete data set is provided in the file \texttt{exact.csv} in \cite{code}.
Our results match up with the values  for $n \in \lbrace 1, \dots, 8\rbrace$ calculated in \cite{Knuth}.
Note that for these values the chosen insertion strategy does not affect the average case (we use the \texttt{left} strategy).

\ifplots
\begin{figure}[ht]
	\vspace{-2mm}
	\begin{minipage}[b]{0.43\textwidth}
		\begin{tikzpicture}
		\begin{axis}[xlabel=$n$,
		ylabel=$\frac{F(n) - n\log n}{n}$,
		ylabel style = {yshift=-7pt, font=\scriptsize},
		xlabel style = {yshift=3pt},
		cycle list name=color list,
		xmode=log,
		xmin=15,
		xmax=150,
		log basis x=2,
		xticklabel style = { font=\scriptsize},
		yticklabel style = { font=\scriptsize},
		width=1.05*\textwidth,
		height=0.6*\axisdefaultheight]
		\addplot table [x expr=\thisrow{n} ,y expr=((\thisrow{avg}-\thisrow{n}*log2(\thisrow{n}))/\thisrow{n}), col sep=tab] {result_exact.csv};
		\end{axis}
		\end{tikzpicture}
		\caption{Computed values of $F(n)$.}
		\label{fig:exactf}
	\end{minipage}%
	\begin{minipage}[b]{0.57\textwidth}
		\footnotesize
		\setlength{\tabcolsep}{0.63mm}
		\begin{tabular}{r|cccccccc}
			$n = $ & 1 & 2 & 3 & 4 & 5 & 6 & 7 & 8 \\
			$\quad\! F(n) \:\!\!\cdot\:\!\! n! = $ & 0 & 2 & 16 & 112 & 832 & 6912 & 62784 & 623232
		\end{tabular}\vspace{2mm}
		\begin{tabular}{r|ccc}
			$n = $ & 9 & 10 & 11 \\
			$\quad\! F(n) \:\!\!\cdot\:\!\! n! = $ & 6743808 & 79292160 & 1013736960
		\end{tabular}\vspace{2mm}
		\begin{tabular}{r|cc}
			$n = $ & 12 & 13\\
			$\quad\! F(n) \:\!\!\cdot\:\!\! n! = $ & 13921182720 & 204489999360
		\end{tabular}\vspace{2mm}
		\begin{tabular}{r|cc}
			$n = $ & 14 & 15\\
			$\quad\! F(n) \:\!\!\cdot\:\!\! n! = $ & 3199119114240 & 53153472153600
		\end{tabular}
		\captionof{table}{Computed values of $F(n)\cdot n!$.}
		\label{tab:exactf}
	\end{minipage}
\end{figure}
\fi

\vspace{-2mm}
\subsubsection*{Improved theoretical upper bounds}
\label{chap:approx}

In this section we improve upon the upper bound from \cite{EdelkampW14} leading to the following result:

\begin{theorem}
	\label{thm:upper_bound}	
		The number of comparisons required in the average case of Merge\-In\-sertion is at most $n \log n - c(x_n)\cdot n \pm \mathcal{O}(\log^2 n)$ where $x_n$ is the fractional part of $\log(3n)$, \ie the unique value in $ \lbrack 0, 1)$ such that $n = 2^{k - \log 3 + x_n}$ for some $k \in \Z$ and	
		$c:\lbrack 0, 1) \to \R$ is given by the following formula: 
		\vspace{-3mm}	\begin{equation*}
		c(x) = (3 - \log 3) - (2 - x - 2^{1-x}) + (1-2^{-x})\left(\frac{3}{2^x+1}-1\right)+\frac{2^{\log 3 - x}}{2292} \ge 1.4005
		\end{equation*}
\end{theorem}
Hence we have obtained a new upper bound for the average case of MergeInsertion which is $n \log n - 1.4005n + \mathcal{O}(\log^2 n)$.
A visual representation of $c(x)$ is provided in \cref{fig:plot_cn}.
The worst case is near $x=0.6$ (\ie $n$ roughly a power of two) where $c(x)$ is just slightly larger than $1.4005$. 

\ifplots
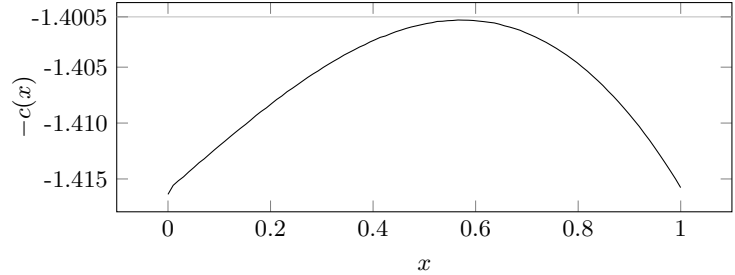
\begin{figure}[t]
	\centering
		\begin{tikzpicture}
		\begin{axis}[xlabel=$x$,
		ylabel=$-c(x)$,
		ytick={-1.405, -1.410, -1.415, -1.420, -1.425},
		yticklabels={-1.405, -1.410, -1.415, -1.420, -1.425},
		extra y ticks={-1.4005},
		extra y tick labels={-1.4005},
		extra y tick style={grid=major},
		width=0.8\textwidth,
		height=0.6*\axisdefaultheight]
		\addplot[domain=0:1,samples=100] {-((3 - (log2(3))) - (2 - x - 2^(1-x)) + (1-2^(-x))*((3)/(2^x+1)-1)+(2^(log2(3) - x))/(2292))};
		\end{axis}
		\end{tikzpicture}
	\caption{Plot of $c(x)$.\vspace{-3mm}}
	\label{fig:plot_cn}
\end{figure}
\fi

The proof of \cref{thm:upper_bound} analyzes the insertion of one batch of elements more carefully than in \cite{EdelkampWW18}.
The exact probability that $b_{t_{k-1}+i}$ is inserted into $j$ elements is given by \cref{th:Yi}.
We are especially interested in the case of $b_{t_{k-1}+u}$ where $u = \lfloor \frac{t_k - t_{k-1}}{2} \rfloor$,
because, if we know $P(Y_u < m)$, then we can use that for all $q < u$ we have $P(Y_q < m) \ge P(Y_u < m)$.

However, the equation from \cref{th:Yi} is hard to work with, so we approximate it with the binomial distribution
$
p(j) = \binom{\left\lceil\frac{u}{2}\right\rceil}{q} (\frac{\lfloor \frac{u}{2} \rfloor}{2t_k-1})^q (\frac{2t_k-1 - \lfloor \frac{u}{2} \rfloor}{2t_k-1})^{\left\lceil\frac{u}{2}\right\rceil - q}
$
with $q=2^k - 1 - j$, that by construction fulfills $\sum_{j=0}^{j_0} p(j) \leq \sum_{j=0}^{j_0} P(Y_u = j) = P(Y_u \le j_0)$ for all $j_0$.
By using the approximation $P(Y_u = j) \approx p(j)$ we can calculate a lower bound for the median of $Y_\frac{t_k - t_{k-1}}{2}$ which is 
$2^k - 1 - \left\lfloor n_B \cdot p_B\right\rfloor \in 2^k - 1 - \frac{2^{k-6}}{3}+\mathcal{O}(1)$.
Thus, with a probability of one half the elements $b_{t_{k-1}+i}$ for $1\leq i \leq u$ are inserted in $\frac{2^{k-6}}{3}$ elements less compared to the worst case.
Combining that with the bounds from \cite{EdelkampWW18} we obtain \Cref{thm:upper_bound}.
The complete proof is given in \Cref{chap:appendix_upper_bound}.

\vspace{-2mm}
\section{Experiments}\label{sec:experiments}
\vspace{-3mm}
In this section we discuss our experiment, which consist of two parts:
first, we evaluate how increasing $t_k$ by some constant factor can reduce the number of comparisons,
then we examine how the combination with the (1,2)-Insertion algorithm as proposed in \cite{12ins} improves MergeInsertion.

We implemented MergeInsertion using a tree based data structure, similar to the Rope data structure\cite{Rope} used in text processing, resulting in a comparably ``fast'' implementation.
Implementation details can be found in \Cref{chap:impl}.
All experiments use the \texttt{left} strategy for binary insertion (see \cref{chap:BinIns}).
The number of comparisons has been averaged over 10 to 10000 runs, depending on the size of the input.

\pgfplotstablesort{\tablefactor}{factor2.csv}

\vspace{-2mm}
\subsubsection*{Increasing $t_k$ by a Constant Factor}

\ifplots
\begin{figure}[t]
	\begin{minipage}[t]{0.35\textwidth}
		\begin{tikzpicture}
		\begin{axis}[xlabel=$n\vphantom{f}$,
		ylabel=$\frac{\text{number of comparisons} - n\log n}{n}$,
		xmode=log,
		log basis x=2,
		xmin=15000,
		xmax=28000,
		ytick={-1.435, -1.430, -1.425, -1.420, -1.415, -1.410},
		yticklabels={-1.435, -1.430, -1.425, -1.420, -1.415, -1.410},
		width=\textwidth,
		height=0.7*\axisdefaultheight]
		\addplot[no markers] table [x=num_elements, y=MI, col sep=tab] {mi.csv};
		\pgfplotstableset{
			col sep=tab,
		};
		\pgfplotstableread{factor.csv}{\mydata};
		\pgfplotstabletranspose[colnames from=factor,input colnames to=n]\otherdatatable{\mydata};
		\addplot[only marks, mark=*] table [x=n, y=1.0] {\otherdatatable};
		\end{axis}
		\end{tikzpicture}
	\caption{$n$ used in \cref{fig:factor}.}
	\label{fig:factor_choice}
\end{minipage}%
\begin{minipage}[t]{0.65\textwidth}
		\begin{tikzpicture}
		\begin{axis}[xlabel=factor $f$,
		ylabel=$\frac{\text{number of comparisons} - n\log n}{n}$,
		legend pos = north west,
		legend style={nodes={scale=0.7, transform shape}},
		cycle list name=color list,
		width=\textwidth,
		height=0.7*\axisdefaultheight]
		\pgfplotsinvokeforeach{16165,17727,18851,19440,20673,21845,22672,23380,24863,26440}{
			\addplot+ table [x=factor, y=#1, col sep=tab] {factor.csv};
			\addlegendentry{\texttt{#1}};
		}

		\end{axis}
		\end{tikzpicture}
	\caption{Effects of replacing $t_k$ with $\hat t_k$.}
	\label{fig:factor}
\end{minipage}
\end{figure}
\fi

\ifplots
\begin{figure}[t]
		\begin{tikzpicture}[spy using outlines={circle, magnification=5, connect spies}]
		\begin{axis}[xlabel=$n$,
		ylabel=$\frac{\text{number of comparisons} - n\log n}{n}$,
		legend pos = north west,
		legend style={nodes={scale=0.85, transform shape}},
		xmode=log,
		log basis x=2,
		cycle list name=color list,
		ytick={-1.435, -1.430, -1.425, -1.420, -1.415, -1.410},
		yticklabels={-1.435, -1.430, -1.425, -1.420, -1.415, -1.410},
		width=\textwidth,
		height=0.75*\axisdefaultheight]
		
		\addplot table [x=num_elements, y=1.0, col sep=tab] {\tablefactor};
		\addlegendentry{$f=1.00$};
		\addplot table [x=num_elements, y=1.02, col sep=tab] {\tablefactor};
		\addlegendentry{$f=1.02$};
		\addplot table [x=num_elements, y=1.03, col sep=tab] {\tablefactor};
		\addlegendentry{$f=1.03$};
		\addplot table [x=num_elements, y=1.04, col sep=tab] {\tablefactor};
		\addlegendentry{$f=1.04$};
		\addplot table [x=num_elements, y=1.05, col sep=tab] {\tablefactor};
		\addlegendentry{$f=1.05$};
		
		\coordinate (spypoint) at (axis cs:87600,-1.4354);
		\coordinate (magnifyglass) at (axis cs:180000,-1.427);
		\coordinate (spypoint2) at (axis cs:524000,-1.419);
		\coordinate (magnifyglass2) at (axis cs:900000,-1.427);
		\end{axis}
		\spy [black, size=2.3cm] on (spypoint)
		in node[fill=white] at (magnifyglass);
		\spy [black, size=2.3cm] on (spypoint2)
		in node[fill=white] at (magnifyglass2);
		\end{tikzpicture}
	\caption{Comparison of different factors $f$ for $\hat t_k$.\vspace{-3mm}}
	\label{fig:factor2}
\end{figure}
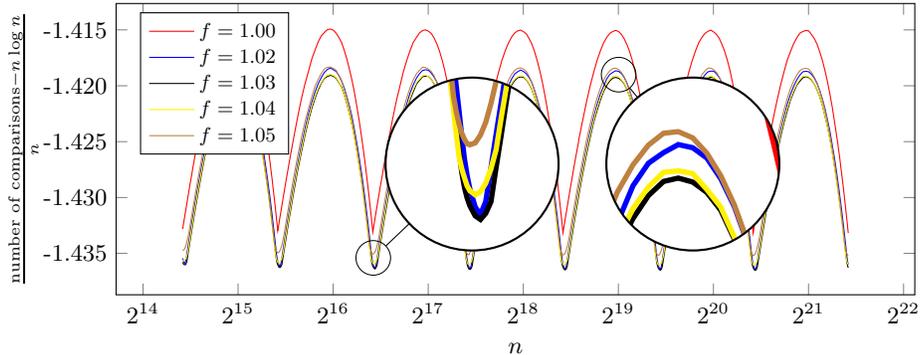
\fi

In this section we modify MergeInsertion by replacing $t_k$ with $\hat t_k = \left\lfloor f\cdot t_k \right\rfloor$~-- otherwise the algorithm is the same.
Originally the numbers $t_k$ have been chosen, such that each element $b_i$ with $t_{k-1} < i \leq t_k$ is inserted into at most $2^k-1$ elements (which is optimal for the worst case).
As we have seen in previous sections many elements are inserted into slightly less than $2^k-1$ elements.
The idea behind increasing $t_k$ by a constant factor $f$ is to allow more elements to be inserted into close to $2^k-1$ elements.

\Cref{fig:factor} shows how different factors $f$ affect the number of comparisons required by MergeInsertion.
The different lines represent different input lengths.
For instance, $n = 21845$ is an input size for which MergeInsertion works best.
An overview of the different input lengths and how original MergeInsertion performs for these can be seen in \Cref{fig:factor_choice}.
The chosen values are assumed to be representative for the entire algorithm.
We observe that for all shown input lengths, multiplying $t_k$ by a factor $f$ between $1.02$ and $1.05$, leads to an improvement.

\Cref{fig:factor2} compares different factors from $1.02$ to $1.05$.
The factor $1.0$ (\ie the original algorithm) is included as a reference.
We observe that all the other factors lead to a considerable improvement compared to $1.0$.
The difference between the factors in the chosen range is rather small.
However, $1.03$ appears to be best out of the tested values.
At $n \approx 2^k/3$ the difference to the information-theoretic lower bound is reduced to $0.007n$, improving upon the original algorithm, which has a difference of $0.01n$ to the optimum.

Another observation we make from \Cref{fig:factor2} is that the plot periodically repeats itself with each power of two.
Thus, we conclude that replacing $t_k$ with $\hat t_k = \left\lfloor f\cdot t_k \right\rfloor$ with $f \in \lbrack 1.02, 1.05 \rbrack$ reduces the number of comparisons required per element by some constant.

\vspace{-2mm}
\subsubsection*{Combination with (1,2)-Insertion}

(1,2)-Insertion is a sorting algorithm presented in \cite{12ins}.
It works by inserting either a single element or two elements at once into an already sorted list.
On its own (1,2)-Insertion is worse than MergeInsertion; however, it can be combined with MergeInsertion.
The combined algorithm works by sorting $m = \max \left\lbrace u_k \mid u_k \le n \right\rbrace$ elements with MergeInsertion.
Then the remaining elements are inserted using (1,2)-Insertion.
Let $u_k = \left\lfloor \left(\frac{4}{3}\right) 2^k \right\rfloor$ denote a point where MergeInsertion is optimal.

In \cref{fig:12ins} we can see that at the point $u_k$ MergeInsertion and the combined algorithm perform the same.
However, in the values following $u_k$ the combined algorithm surpasses MergeInsertion until at one point close to the next optimum MergeInsertion is better once again.
In their paper Iwama and Teruyama calculated that for $0.638 \le \frac{n}{2^{\lceil\log n\rceil}} \le \frac{2}{3}$ MergeInsertion is better than the combined algorithm.
The fraction $\frac{2}{3}$ corresponds to the point where MergeInsertion is optimal.
They derived the constant $0.638$ from their theoretical analysis using the upper bound for MergeInsertion from \cite{EdelkampW14}.
Comparing this to our experimental results we observe that the range where MergeInsertion is better than the combined algorithm starts at $n\approx 2^{17.242}$.
This yields $\frac{2^{17.242}}{2^{18}} = 2^{17.242-18} = 2^{-0.758} \approx 0.591$.
Hence the range where MergeInsertion is better than the combined algorithm is $0.591 \le \frac{n}{2^{\lceil\log n\rceil}} \le \frac{2}{3}$, which is slightly larger than the theoretical analysis suggested.
Also shown in \cref{fig:12ins} is the combined algorithm where we additionally apply our suggestion of replacing $t_k$ by $\hat t_k  = \left\lfloor f\cdot t_k \right\rfloor$ with $f=1.03$. This leads to an additional improvement and comes even closer to the lower bound of $\log(n!)$.

\ifplots
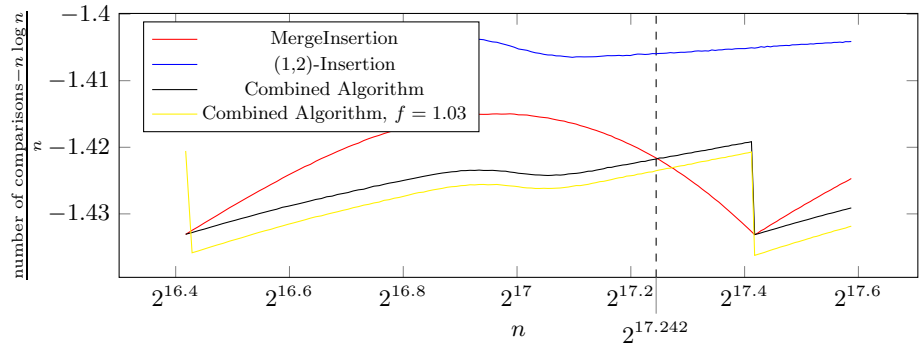
\begin{figure}[t]
		\begin{tikzpicture}
		\begin{axis}[xlabel=$n$,
		ylabel=$\frac{\text{number of comparisons} - n\log n}{n}$,
		cycle list name=color list,
		legend pos = north west,
		legend style={nodes={scale=0.8, transform shape}},
		xmode=log,
		log basis x=2,
		extra x ticks={155000},
		extra x tick labels={$2^{17.242}$},
		extra tick style={
			tick align=outside,
			tick pos=left,
			grid style={dashed,black},
			grid=major
		},
		extra x tick style={
			major tick length=1.25\baselineskip
		},
		extra y tick style={
			major tick length=2.5em
		},
		width=\textwidth,
		height=0.7*\axisdefaultheight]
		\addplot table [x=num_elements, y=MI, col sep=tab] {experiments2.csv};
		\addlegendentry{MergeInsertion}
		\addplot table [x=num_elements, y=12InsertionImproved, col sep=tab] {experiments2.csv};
		\addlegendentry{(1,2)-Insertion}
		\addplot table [x=num_elements, y=MIw12Ins, col sep=tab] {experiments2.csv};
		\addlegendentry{Combined Algorithm}
		\addplot table [x=num_elements, y=12InsertionCombinedfactor103, col sep=tab] {experiments4.csv};
		\addlegendentry{Combined Algorithm, $f=1.03$}
		\end{axis}
		\end{tikzpicture}
	\caption{Experimental results comparing MergeInsertion, (1,2)-Insertion and the combined algorithm.\vspace{-3mm}}
	\label{fig:12ins}
\end{figure}
\fi

\vspace{-2mm}
\subsubsection*{Conclusion and Outlook}

We improved the previous upper bound of $n \log n - 1.3999n + o(n)$ to $n \log n - 1.4005n + o(n)$ for the average number of comparisons of MergeInsertion. However, there still is a gap between the number of comparisons required by MergeInsertion and this upper bound.

In \cref{chap:approx} we used a binomial distribution to approximate the probability of an element being inserted into a specific number of elements during the insertion step.
However, the difference between our approximation and the actual probability distribution is rather large.
Finding an approximation which reduces that gap while still being simple to analyze with respect to its mean would facilitate further improvements to the upper bound.

Our suggestion of increasing $t_k$ by a constant factor $f$ reduced the number of comparisons required per element by some constant. However, we do not have a proof for this.
Thus, future research could try to determine the optimal value for the factor $f$ as well as to study how this suggestion affects the worst-case.

\clearpage
\appendix
\section{Tables and Figures}

\begin{figure}[ht]
	\centering
	\begin{subfigure}[b]{\textwidth/3}
		\begin{center}
			\begin{tikzpicture}
			\foreach \x in {2, 3}
			\filldraw(\x - 0.15,0) circle (2pt) node [below] {$b_{\x}$}
			(\x, 0.8) circle (2pt) node [above] {$a_{\x}$};
			\foreach \x in {2, 3}
			\draw[-latex,shorten >=2pt] (\x - 0.15,0) -- (\x, 0.8);
			\filldraw (0, 0.8) circle (2pt) node [above] {$x_1$};
			\filldraw (1, 0.8) circle (2pt) node [above] {$x_2$};
			\foreach \x in {0, 1, ..., 2}
			\draw[-latex,shorten >=2pt] (\x, 0.8) -- (\x + 1, 0.8);
			\end{tikzpicture}
		\end{center}
		\caption{Batch $k=2$}
		\label{fig:batches2}
	\end{subfigure}
	~
	\begin{subfigure}[b]{\textwidth/2}
		\begin{center}
			\begin{tikzpicture}
			\foreach \x in {4, 5}
			\filldraw(\x - 0.15,0) circle (2pt) node [below] {$b_{\x}$}
			(\x, 0.8) circle (2pt) node [above] {$a_{\x}$};
			\foreach \x in {4, 5}
			\draw[-latex,shorten >=2pt] (\x - 0.15,0) -- (\x, 0.8);
			\filldraw (-2, 0.8) circle (2pt) node [above] {$x_1$};
			\filldraw (-1, 0.8) circle (2pt) node [above] {$x_2$};
			\filldraw (0, 0.8) circle (2pt) node [above] {$x_3$};
			\filldraw (1, 0.8) circle (2pt) node [above] {$x_4$};
			\filldraw (2, 0.8) circle (2pt) node [above] {$x_5$};
			\filldraw (3, 0.8) circle (2pt) node [above] {$x_6$};
			\foreach \x in {-2, -1, ..., 4}
			\draw[-latex,shorten >=2pt] (\x, 0.8) -- (\x + 1, 0.8);
			\end{tikzpicture}
		\end{center}
		\caption{Batch $k=3$}
		\label{fig:batches3}
	\end{subfigure}

	\begin{subfigure}[b]{\textwidth}
		\begin{center}
			\begin{tikzpicture}
			\foreach \x in {6, 7, ..., 11}
			\filldraw(\x - 0.15,0) circle (2pt) node [below] {$b_{\x}$}
			(\x, 0.8) circle (2pt) node [above] {$a_{\x}$};
			\foreach \x in {6, 7, ..., 11}
			\draw[-latex,shorten >=2pt](\x - 0.15,0) -- (\x, 0.8);
			\filldraw (0.5, 0.8) circle (2pt) node [above] {$x_1$};
			\filldraw (1.5, 0.8) circle (2pt) node [above] {$x_2$};
			\filldraw (2.5, 0.8) circle (2pt) node [above] {$x_3$};
			\filldraw (4, 0.8) circle (2pt) node [above] {$x_9$};
			\filldraw (5, 0.8) circle (2pt) node [above] {$x_{10}$};
			\foreach \x in {0.5, 1.5}
			\draw[-latex,shorten >=2pt] (\x, 0.8) -- (\x + 1, 0.8);
			\foreach \x in {4, 5, ..., 10}
			\draw[-latex,shorten >=2pt] (\x, 0.8) -- (\x + 1, 0.8);
			\draw[-latex,shorten >=2pt,dashed] (2.5, 0.8) -- (4, 0.8);
			\end{tikzpicture}
		\end{center}
		\caption{Batch $k=4$}
		\label{fig:batches4}
	\end{subfigure}

	\caption{Batches of the elements $b_{t_k}$ to $b_{t_{k-1}+1}$ for $k \in \{2, 3, 4\}$}\label{fig:batches}
\end{figure}
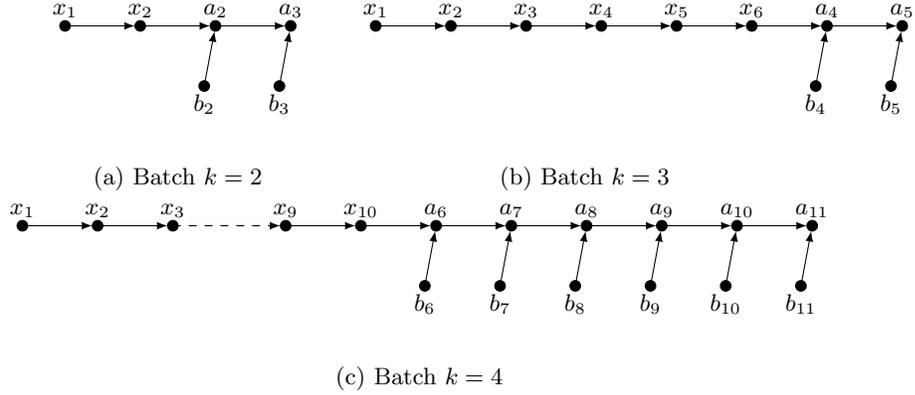

\begin{table}
	\resizebox{\textwidth}{!}{
		\begin{tabular}{|l|r|r|r|r|r|r|}
			\hline
			$i$ & 1 & 2 & 3 & 4 & 5 & 6 \\
			\hline
			$P(X_i = 0)$ & $\frac{1}{11}$ & $\frac{1}{11}\cdot\frac{12}{13}$ & $\frac{1}{11}\cdot\frac{12}{13}\cdot\frac{14}{15}$ & $\frac{1}{11}\cdot\frac{12}{13}\cdot\frac{14}{15}\cdot\frac{16}{17}$ & $\frac{1}{11}\cdot\frac{12}{13}\cdot\frac{14}{15}\cdots\frac{18}{19}$ & $\frac{1}{11}\cdot\frac{12}{13}\cdot\frac{14}{15}\cdots\frac{20}{21}$ \\
			\hline
			$P(X_i = 1)$ & $\frac{1}{11}$ & $\frac{1}{11}\cdot\frac{12}{13}$ & $\frac{1}{11}\cdot\frac{12}{13}\cdot\frac{14}{15}$ & $\frac{1}{11}\cdot\frac{12}{13}\cdot\frac{14}{15}\cdot\frac{16}{17}$ & $\frac{1}{11}\cdot\frac{12}{13}\cdot\frac{14}{15}\cdots\frac{18}{19}$ & $\frac{1}{11}\cdot\frac{12}{13}\cdot\frac{14}{15}\cdots\frac{20}{21}$ \\
			\hline
			$P(X_i = 2)$ & $\frac{1}{11}$ & $\frac{1}{11}\cdot\frac{12}{13}$ & $\frac{1}{11}\cdot\frac{12}{13}\cdot\frac{14}{15}$ & $\frac{1}{11}\cdot\frac{12}{13}\cdot\frac{14}{15}\cdot\frac{16}{17}$ & $\frac{1}{11}\cdot\frac{12}{13}\cdot\frac{14}{15}\cdots\frac{18}{19}$ & $\frac{1}{11}\cdot\frac{12}{13}\cdot\frac{14}{15}\cdots\frac{20}{21}$ \\
			\hline
			$P(X_i = 3)$ & $\frac{1}{11}$ & $\frac{1}{11}\cdot\frac{12}{13}$ & $\frac{1}{11}\cdot\frac{12}{13}\cdot\frac{14}{15}$ & $\frac{1}{11}\cdot\frac{12}{13}\cdot\frac{14}{15}\cdot\frac{16}{17}$ & $\frac{1}{11}\cdot\frac{12}{13}\cdot\frac{14}{15}\cdots\frac{18}{19}$ & $\frac{1}{11}\cdot\frac{12}{13}\cdot\frac{14}{15}\cdots\frac{20}{21}$ \\
			\hline
			$P(X_i = 4)$ & $\frac{1}{11}$ & $\frac{1}{11}\cdot\frac{12}{13}$ & $\frac{1}{11}\cdot\frac{12}{13}\cdot\frac{14}{15}$ & $\frac{1}{11}\cdot\frac{12}{13}\cdot\frac{14}{15}\cdot\frac{16}{17}$ & $\frac{1}{11}\cdot\frac{12}{13}\cdot\frac{14}{15}\cdots\frac{18}{19}$ & $\frac{1}{11}\cdot\frac{12}{13}\cdot\frac{14}{15}\cdots\frac{20}{21}$ \\
			\hline
			$P(X_i = 5)$ & $\frac{1}{11}$ & $\frac{1}{11}\cdot\frac{12}{13}$ & $\frac{1}{11}\cdot\frac{12}{13}\cdot\frac{14}{15}$ & $\frac{1}{11}\cdot\frac{12}{13}\cdot\frac{14}{15}\cdot\frac{16}{17}$ & $\frac{1}{11}\cdot\frac{12}{13}\cdot\frac{14}{15}\cdots\frac{18}{19}$ & $\frac{1}{11}\cdot\frac{12}{13}\cdot\frac{14}{15}\cdots\frac{20}{21}$ \\
			\hline
			$P(X_i = 6)$ & $\frac{1}{11}$ & $\frac{1}{11}\cdot\frac{12}{13}$ & $\frac{1}{11}\cdot\frac{12}{13}\cdot\frac{14}{15}$ & $\frac{1}{11}\cdot\frac{12}{13}\cdot\frac{14}{15}\cdot\frac{16}{17}$ & $\frac{1}{11}\cdot\frac{12}{13}\cdot\frac{14}{15}\cdots\frac{18}{19}$ & $\frac{1}{11}\cdot\frac{12}{13}\cdot\frac{14}{15}\cdots\frac{20}{21}$ \\
			\hline
			$P(X_i = 7)$ & $\frac{1}{11}$ & $\frac{1}{11}\cdot\frac{12}{13}$ & $\frac{1}{11}\cdot\frac{12}{13}\cdot\frac{14}{15}$ & $\frac{1}{11}\cdot\frac{12}{13}\cdot\frac{14}{15}\cdot\frac{16}{17}$ & $\frac{1}{11}\cdot\frac{12}{13}\cdot\frac{14}{15}\cdots\frac{18}{19}$ & $\frac{1}{11}\cdot\frac{12}{13}\cdot\frac{14}{15}\cdots\frac{20}{21}$ \\
			\hline
			$P(X_i = 8)$ & $\frac{1}{11}$ & $\frac{1}{11}\cdot\frac{12}{13}$ & $\frac{1}{11}\cdot\frac{12}{13}\cdot\frac{14}{15}$ & $\frac{1}{11}\cdot\frac{12}{13}\cdot\frac{14}{15}\cdot\frac{16}{17}$ & $\frac{1}{11}\cdot\frac{12}{13}\cdot\frac{14}{15}\cdots\frac{18}{19}$ & $\frac{1}{11}\cdot\frac{12}{13}\cdot\frac{14}{15}\cdots\frac{20}{21}$ \\
			\hline
			$P(X_i = 9)$ & $\frac{1}{11}$ & $\frac{1}{11}\cdot\frac{12}{13}$ & $\frac{1}{11}\cdot\frac{12}{13}\cdot\frac{14}{15}$ & $\frac{1}{11}\cdot\frac{12}{13}\cdot\frac{14}{15}\cdot\frac{16}{17}$ & $\frac{1}{11}\cdot\frac{12}{13}\cdot\frac{14}{15}\cdots\frac{18}{19}$ & $\frac{1}{11}\cdot\frac{12}{13}\cdot\frac{14}{15}\cdots\frac{20}{21}$ \\
			\hline
			$P(X_i = 10)$ & $\frac{1}{11}$ & $\frac{1}{11}\cdot\frac{12}{13}$ & $\frac{1}{11}\cdot\frac{12}{13}\cdot\frac{14}{15}$ & $\frac{1}{11}\cdot\frac{12}{13}\cdot\frac{14}{15}\cdot\frac{16}{17}$ & $\frac{1}{11}\cdot\frac{12}{13}\cdot\frac{14}{15}\cdots\frac{18}{19}$ & $\frac{1}{11}\cdot\frac{12}{13}\cdot\frac{14}{15}\cdots\frac{20}{21}$ \\
			\hline
			$P(X_i = 11)$ & $0$ & $\frac{1}{13}$ & $\frac{1}{13}\cdot\frac{14}{15}$ & $\frac{1}{13}\cdot\frac{14}{15}\cdot\frac{16}{17}$ & $\frac{1}{13}\cdot\frac{14}{15}\cdot\frac{16}{17}\cdot\frac{18}{19}$ & $\frac{1}{13}\cdot\frac{14}{15}\cdot\frac{16}{17}\cdots\frac{20}{21}$ \\
			\hline
			$P(X_i = 12)$ & $0$ & $0$ & $\frac{1}{15}$ & $\frac{1}{15}\cdot\frac{16}{17}$ & $\frac{1}{15}\cdot\frac{16}{17}\cdot\frac{18}{19}$ & $\frac{1}{15}\cdot\frac{16}{17}\cdot\frac{18}{19}\cdot\frac{20}{21}$ \\
			\hline
			$P(X_i = 13)$ & $0$ & $0$ & $0$ & $\frac{1}{17}$ & $\frac{1}{17}\cdot\frac{18}{19}$ & $\frac{1}{17}\cdot\frac{18}{19}\cdot\frac{20}{21}$ \\
			\hline
			$P(X_i = 14)$ & $0$ & $0$ & $0$ & $0$ & $\frac{1}{19}$ & $\frac{1}{19}\cdot\frac{20}{21}$ \\
			\hline
			$P(X_i = 15)$ & $0$ & $0$ & $0$ & $0$ & $0$ & $\frac{1}{21}$ \\
			\hline
	\end{tabular}}
	\vspace{5mm}
	\caption{Values of $P(X_i = j)$ for $k = 4$.}
	\label{tab:insert}
\end{table}

\ifplots
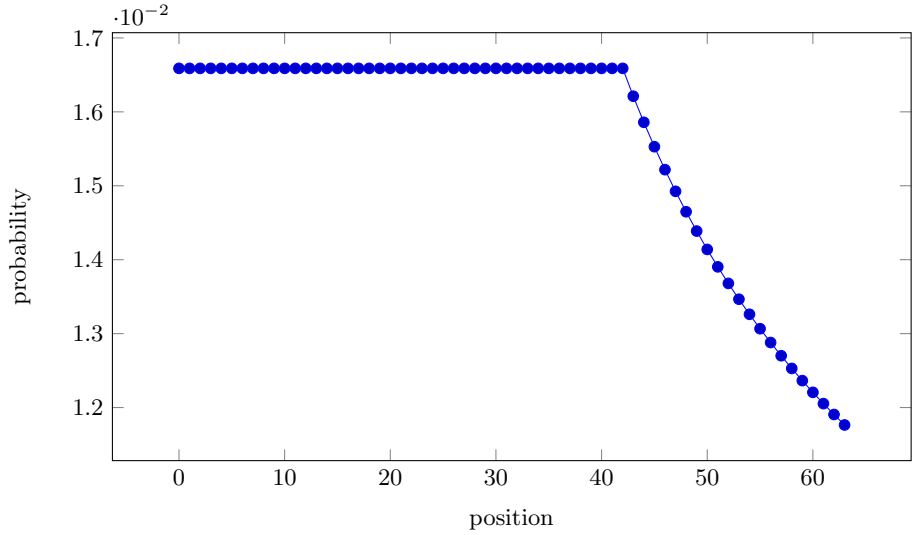
\begin{figure}[h]
	\begin{center}
		\begin{tikzpicture}
		\begin{axis}[xlabel=position,
		ylabel=probability,
		width=\textwidth,
		height=\axisdefaultheight]
		\addplot table [x=i, y=P, col sep=tab] {distr_insert.csv};
		\end{axis}
		\end{tikzpicture}
	\end{center}
	\caption{Probabilities of different positions when inserting $b_{t_k}$ where $k=6$.}
	\label{fig:prob_ins}
\end{figure}

\begin{figure}[h]
	\begin{center}
		\begin{tikzpicture}
		\begin{axis}[xlabel=$n$,
		cycle list name=color list,
		legend pos = north west,
		xmode=log,
		log basis x=2,
		width=\textwidth,
		height=\axisdefaultheight]
		\addplot table [x=num_elements, y=1.0, col sep=tab] {\tablefactor};
		\addlegendentry{Experimental result};
		\addplot[domain=22000:2800000,samples=500] {-((3 - (log10(3)/log10(2))) - (2 - (log2(3*x)-floor(log2(3*x))) - 2^(1-(log2(3*x)-floor(log2(3*x))))) + (1-2^(-(log2(3*x)-floor(log2(3*x)))))*((3)/(2^(log2(3*x)-floor(log2(3*x)))+1)-1)+(2^(log10(3)/log10(2) - (log2(3*x)-floor(log2(3*x)))))/(2292))};
		\addlegendentry{Upper Bound};
		\end{axis}
		\end{tikzpicture}
	\end{center}
	\caption{\small Comparing experimental results with the upper bound from \cref{thm:upper_bound}.}
	\label{fig:impl}
\end{figure}
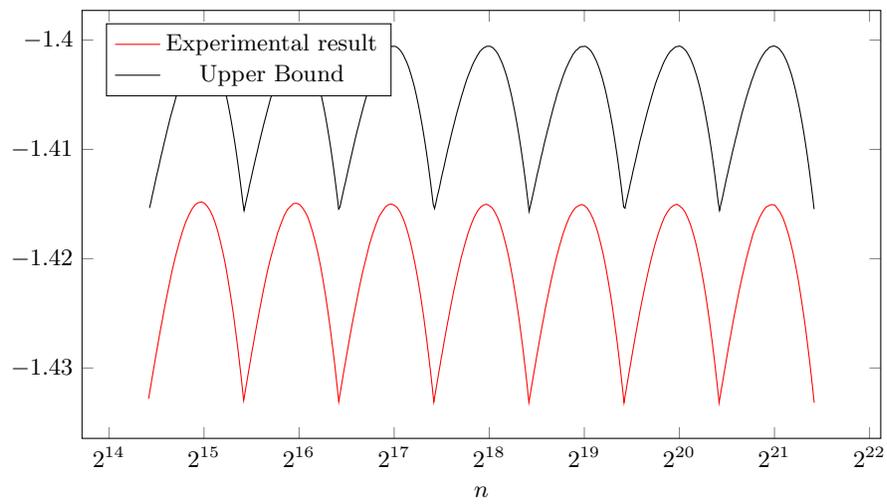
\fi

\begin{algorithm}
	\caption{Binary Insertion}
	\small	\label{alg:binins}
	\begin{algorithmic}[1]
		\Procedure{Insert}{$a$,$x_1$, ..., $x_n$}
		\If{$n=0$}
		\State \Return $a$
		\EndIf
		\State $k \gets \lfloor \log n \rfloor$
		\State $c \gets
		\begin{cases}
		\left\lfloor\frac{n+1}{2}\right\rfloor & \text{strategy \texttt{center-left}}\\
		\left\lceil\frac{n+1}{2}\right\rceil & \text{strategy \texttt{center-right}}\\
		\max\lbrace n-2^k+1, 2^{k-1}\rbrace & \text{strategy \texttt{left}}\\
		\min\lbrace 2^k, n - 2^{k-1}+1\rbrace & \text{strategy \texttt{right}}\\
		\end{cases}$
		\If{$a < x_c$}
		\State $y_1, ..., y_c \gets $\Call{Insert}{$a$,$x_1$, ..., $x_{c-1}$}
		\State \Return $y_1, ..., y_c, x_c, ..., x_n$
		\Else
		\State $y_c, ..., y_n \gets $\Call{Insert}{$a$,$x_{c+1}$, ..., $x_{n}$}
		\State \Return $x_1, ..., x_c, y_c, ..., y_n$
		\EndIf
		\EndProcedure
	\end{algorithmic}
\end{algorithm}

\clearpage
\section{Missing Proofs}

\subsection{Proof of \Cref{th:Xi}}
\label{proof:Xi}
For an arbitrary $k$ we can calculate the probabilities $P(X_i = j)$ with the following recursive scheme.
We start with $P(X_1 = j)$.
This corresponds to the insertion of $b_{t_{k-1}+1}$ into $x_1,\dots,x_{2t_{k-1}}$.
The probability of all those is uniformly distributed, so $P(X_1 = j) = \frac{1}{2t_{k-1} + 1}$ for $0 \le j \le 2t_{k-1}$.

For $i>1$ we can  express $P(X_i = j)$ in terms of $P(X_{i-1} = j)$.
Observe that when inserting $b_{t_{k-1}+i}$ there are $2t_{k-1}+2i-2$ elements known to be smaller than $a_{t_{k-1}+i}$.
These are $x_1,\dots,x_{2t_{k-1}}$ and $a_{t_{k-1}+1},\dots,a_{t_{k-1}+i-1}$ as well as the corresponding $b$'s.
The number of elements known to be smaller than $a_{t_{k-1}+i-1}$ is one less: just $2t_{k-1}+2i-3$.
As a result the probability that $b_{t_{k-1}+i}$ is inserted between $a_{t_{k-1}+i-1}$ and $a_{t_{k-1}+i}$ is $P(X_i = 2t_{k-1}+i-1)=\frac{1}{2t_{k-1}+2i-1}$.
The probability that is ends up in one of the other positions consequently is $P(0 \le X_i < 2t_{k-1}+i-1)=\frac{2t_{k-1}+2i-2}{2t_{k-1}+2i-1}$.
If we know that $b_{t_{k-1}+i}$ is inserted into one of those other positions, then it is inserted into exactly the same elements as $b_{t_{k-1}+i-1}$, thus we can write $P(X_i = j)=\frac{2t_{k-1}+2i-2}{2t_{k-1}+2i-1}P(X_{i-1}=j)$.
This leads to \cref{eq:Xieqj_ugly}.

\begin{equation}
\resizebox{\textwidth}{!}{$
	P(X_i = j) =  \left\{\begin{array}{ll}
	\displaystyle\left(\prod_{l=1}^{i-1}2t_{k-1} + 2l\right) \cdot \left(\prod_{l=1}^{i}2t_{k-1} + 2l - 1\right)^{-1} & \text{if } 0 \leq j \leq 2t_{k-1}  \\
	\displaystyle\left(\prod_{l=j-2t_{k-1}+1}^{i-1}2t_{k-1} + 2l\right) \cdot \left(\prod_{l=j-2t_{k-1}+1}^{i}2t_{k-1} + 2l - 1\right)^{-1} & \text{if } 2t_{k-1} < j < 2t_{k-1} + i \\
	0 & \text{otherwise.}
	\end{array}\right.$}
\label{eq:Xieqj_ugly}
\end{equation}

It remains to simplify \cref{eq:Xieqj_ugly}.
We begin with the first case:

\begin{equation}
\begin{split}
& \left(\prod_{l=1}^{i-1}2t_{k-1} + 2l\right) \cdot \left(\prod_{l=1}^{i}2t_{k-1} + 2l - 1\right)^{-1}  \\
&= \left(\prod_{l = t_{k-1} + 1}^{t_{k-1} + i - 1} 2l\right) \cdot \left(\prod_{l=2t_{k-1} + 1}^{2t_{k-1} + 2i - 1}l\right)^{-1} \cdot \left(\prod_{l=t_{k-1} + 1}^{t_{k-1} + i -1} 2l\right) \\
&= \left(\prod_{l=1}^{t_{k-1} + i -1} 2l\right) \cdot \left(\prod_{l=1}^{t_{k-1}}2l\right)^{-1} \cdot \left(\prod_{l=1}^{2t_{k-1}+2i-1}l\right)^{-1}  \\&\qquad\cdot \left(\prod_{l=1}^{2t_{k-1}}l\right) \cdot \left(\prod_{l=1}^{t_{k-1} +i - 1}2l\right) \cdot \left(\prod_{l=1}^{t_{k-1}}2l\right)^{-1}  \\
&= 2^{2i-2} \left(\frac{\left(t_{k-1}+i-1\right)!}{\left(t_{k-1}\right)!}\right)^2 \frac{\left(2t_{k-1}\right)!}{\left(2t_{k-1}+2i-1\right)!}
\end{split}
\label{eq:Xieqj_first}
\end{equation}

For the second case we have

\begin{equation}
\begin{split}
& \left(\prod_{l=j-2t_{k-1}+1}^{i-1}2t_{k-1} + 2l\right) \cdot \left(\prod_{l=j-2t_{k-1}+1}^{i}(2t_{k-1} + 2l - 1\right)^{-1}  \\
&= \left(\prod_{l = j - t_{k-1} + 1}^{t_{k-1} + i - 1} 2l\right) \cdot \left(\prod_{l=2j - 2t_{k-1} + 1}^{2t_{k-1} + 2i - 1}l\right)^{-1} \cdot \left(\prod_{l=j - t_{k-1} + 1}^{t_{k-1} + i -1} 2l\right) \\
&= \left(\prod_{l=1}^{t_{k-1} + i -1} 2l\right) \cdot \left(\prod_{l=1}^{j-t_{k-1}}2l\right)^{-1} \cdot \left(\prod_{l=1}^{2t_{k-1}+2i-1}l\right)^{-1} \\&\qquad\cdot \left(\prod_{l=1}^{2j-2t_{k-1}}l\right) \cdot \left(\prod_{l=1}^{t_{k-1} +i - 1}2l\right) \cdot \left(\prod_{l=1}^{j-t_{k-1}}2l\right)^{-1}  \\
&= 2^{4t_{k-1}-2j+2i-2} \left(\frac{\left(t_{k-1}+i-1\right)!}{\left(j-t_{k-1}\right)!}\right)^2 \frac{\left(2j-2t_{k-1}\right)!}{\left(2t_{k-1}+2i-1\right)!}
\end{split}
\label{eq:Xieqj_second}
\end{equation}

By substitution of (\ref{eq:Xieqj_first}) and (\ref{eq:Xieqj_second}) in (\ref{eq:Xieqj_ugly}) we obtain \cref{th:Xi}.

\subsection{Proof of \Cref{th:Yi}}
\label{proof:Yi}

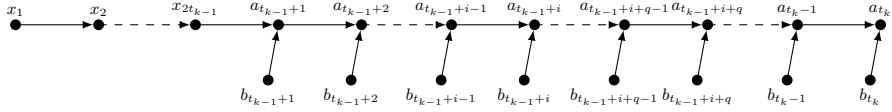
\begin{figure}
	\centering
	\begin{tikzpicture}[scale=0.92]
	\filldraw (0.6, 0.8) circle (2pt) node [above] {\scalebox{0.7}{$x_1$}}
	(1.8, 0.8) circle (2pt) node [above] {\scalebox{0.7}{$x_2$}}
	(3.2, 0.8) circle (2pt) node [above] {\scalebox{0.7}{$x_{2t_{k-1}}$}}
	(4.4, 0.8) circle (2pt) node [above] {\scalebox{0.7}{$a_{t_{k-1}+1}$}}
	(5.6, 0.8) circle (2pt) node [above] {\scalebox{0.7}{$a_{t_{k-1}+2}$}}
	(6.9, 0.8) circle (2pt) node [above] {\scalebox{0.7}{$a_{t_{k-1}+i-1}$}}
	(8.1, 0.8) circle (2pt) node [above] {\scalebox{0.7}{$a_{t_{k-1}+i}$}}
	(9.4, 0.8) circle (2pt) node [above] {\scalebox{0.7}{$a_{t_{k-1}+i+q-1}$}}
	(10.6, 0.8) circle (2pt) node [above] {\scalebox{0.7}{$a_{t_{k-1}+i+q}$}}
	(11.9, 0.8) circle (2pt) node [above] {\scalebox{0.7}{$a_{t_k-1}$}}
	(13.1, 0.8) circle (2pt) node [above] {\scalebox{0.7}{$a_{t_k}$}};
	
	\draw[-latex,shorten >=2pt](0.6, 0.8) -- (1.8, 0.8);
	\draw[-latex,shorten >=2pt](3.2, 0.8) -- (4.4, 0.8);
	\draw[-latex,shorten >=2pt](4.4, 0.8) -- (5.6, 0.8);
	\draw[-latex,shorten >=2pt](6.9, 0.8) -- (8.1, 0.8);
	\draw[-latex,shorten >=2pt](9.4, 0.8) -- (10.6, 0.8);
	\draw[-latex,shorten >=2pt](11.9, 0.8) -- (13.1, 0.8);
	\draw[-latex,shorten >=2pt,dashed] (1.8, 0.8) -- (3.2, 0.8);
	\draw[-latex,shorten >=2pt,dashed] (5.6, 0.8) -- (6.9, 0.8);
	\draw[-latex,shorten >=2pt,dashed] (8.1, 0.8) -- (9.4, 0.8);
	\draw[-latex,shorten >=2pt,dashed] (10.6, 0.8) -- (11.9, 0.8);
	\filldraw(4.4 - 0.15,0) circle (2pt) node [below] {\scalebox{0.7}{$b_{t_{k-1}+1}$}};
	\filldraw(5.6 - 0.15,0) circle (2pt) node [below] {\scalebox{0.7}{$b_{t_{k-1}+2}$}};
	\filldraw(6.9 - 0.15,0) circle (2pt) node [below] {\scalebox{0.7}{$b_{t_{k-1}+i-1}$}};
	\filldraw(8.1 - 0.15,0) circle (2pt) node [below] {\scalebox{0.7}{$b_{t_{k-1}+i}$}};
	\filldraw(9.4 - 0.15,0) circle (2pt) node [below] {\scalebox{0.7}{$b_{t_{k-1}+i+q-1}$}};
	\filldraw(10.6 - 0.15,0) circle (2pt) node [below] {\scalebox{0.7}{$b_{t_{k-1}+i+q}$}};
	\filldraw(11.9 - 0.15,0) circle (2pt) node [below] {\scalebox{0.7}{$b_{t_k-1}$}};
	\filldraw(13.1 - 0.15,0) circle (2pt) node [below] {\scalebox{0.7}{$b_{t_k}$}};
	\draw[-latex,shorten >=2pt](4.4 - 0.15,0) -- (4.4, 0.8);
	\draw[-latex,shorten >=2pt](5.6 - 0.15,0) -- (5.6, 0.8);
	\draw[-latex,shorten >=2pt](6.9 - 0.15,0) -- (6.9, 0.8);
	\draw[-latex,shorten >=2pt](8.1 - 0.15,0) -- (8.1, 0.8);
	\draw[-latex,shorten >=2pt](9.4 - 0.15,0) -- (9.4, 0.8);
	\draw[-latex,shorten >=2pt](10.6 - 0.15,0) -- (10.6, 0.8);
	\draw[-latex,shorten >=2pt](11.9 - 0.15,0) -- (11.9, 0.8);
	\draw[-latex,shorten >=2pt](13.1 - 0.15,0) -- (13.1, 0.8);
	\end{tikzpicture}
	\caption{Configuration where one batch of $t_k-t_{k-1}$ elements remains to be inserted. The elements $b_{t_{k-1}+i}$ and $b_{t_{k-1}+i+q}$ are drawn.}
	\label{fig:proof_px_cond}
\end{figure}

Recall the definitions of $Y_i$, $\tilde{Y}_{i,q}$ and their relation:

\begin{equation}
Y_i: \omega \mapsto \left\vert \left\{ v \in \oneset{x_1, \dots, x_{2^k} } \cup \{ b_{t_{k-1}+i+ 1}, \dots, b_{t_{k}}\}  \mid \omega(v) < \omega(a_{t_{k-1}+i}) \right\} \right\vert
\end{equation}

\begin{equation}
\tilde{Y}_{i,q}: \omega \mapsto \left\vert \left\{  v \in \{b_{t_{k-1}+i+1}, \dots, b_{t_{k-1}+i+q}\}\mid \omega(v) < \omega(a_{t_{k-1}+i})  \right\} \right\vert
\end{equation}

\begin{equation}
Y_i = \tilde{Y}_{i,t_k-t_{k-1}-i} + 2t_{k-1} + i - 1
\label{eq:relYiYiq}
\end{equation}

To proof \cref{th:Yi} we start with the following closed form for the probability $P(\tilde{Y}_{i,q} = j)$.\footnote{The first part of \cref{eq:Yiq_closed}: $\frac{(2q-j)!}{2^{q-j}j!(q-j)!}$, when substituting $q=n$ and $j=n-k$ yields\\ $a(n,k)=\frac{(n+k)!}{2^{k}(n-k)!k!}$ which is the number sequence A001498 from The On-Line Encyclopedia of Integer Sequences \url{https://oeis.org/A001498}.} 
\begin{equation}
P(\tilde{Y}_{i,q} = j) = \frac{(2q-j)!}{2^{q-j}j!(q-j)!}2^q\frac{(2t_{k-1}+2i+j-1)!}{(2t_{k-1}+2i+2q-1)!}\frac{(t_{k-1}+i+q-1)!}{(t_{k-1}+i-1)!}
\label{eq:Yiq_closed}
\end{equation}

From the definition of $\tilde{Y}_{i,q}$ we can see that $0 \le \tilde{Y}_{i,q} \le q$ thus $P(\tilde{Y}_{i,0} = 0) = 1$.
This also holds for \cref{eq:Yiq_closed}.
\begin{equation}
P(\tilde{Y}_{i,0} = 0) = \frac{0!}{2^0 \cdot 0! \cdot 0!}2^0\frac{(2t_{k-1} +2i - 1)!}{(2t_{k-1} +2i - 1)!}\frac{(t_{k-1}+i-1)!}{(t_{k-1}+i-1)!} = 1
\end{equation}

Recall that for $q>0$ there are two possibilities:
\begin{enumerate}
	\item $\tilde{Y}_{i,q-1} = j-1$ and $X_{i+q} < 2t_{k-1}+i$. Informally speaking that means out of $\{b_{t_{k-1}+i+1}, \dots, b_{t_{k-1}+i+q-1}\}$ there have been $j-1$ elements inserted before $a_{t_{k-1}+i}$ and $b_{t_{k-1}+i+q}$ is inserted before $a_{t_{k-1}+i}$.
	\item $\tilde{Y}_{i,q-1} = j$ and $X_{i+q} \ge 2t_{k-1}+i$. Informally speaking that means out of $\{b_{t_{k-1}+i+1}, \dots, b_{t_{k-1}+i+q-1}\}$ there have been $j$ elements inserted before $a_{t_{k-1}+i}$ and $b_{t_{k-1}+i+q}$ is inserted after $a_{t_{k-1}+i}$.
\end{enumerate}
Note that the first case requires $j > 0$ and the second case requires $j < q$ so we look at $j=0$ and $j=q$ separately.

Using Bayes' theorem we obtain the following identities:

\begin{equation}
\resizebox{\textwidth}{!}{$
\begin{split}
P(X_{i+q} \ge 2t_{k-1} + i \land \tilde{Y}_{i,q-1} = 0)&=P(X_{i+q} \ge 2t_{k-1} + i \:\vert\: \tilde{Y}_{i,q-1} = 0) \cdot P(\tilde{Y}_{i,q-1} = 0)\\
P(X_{i+q} < 2t_{k-1}+i \land \tilde{Y}_{i,q-1}=q-1)
&=P(X_{i+q} < 2t_{k-1}+i \:\vert\: \tilde{Y}_{i,q-1}=q-1)\cdot P(\tilde{Y}_{i,q-1}=q-1)\\
\end{split}$}
\end{equation}

The probability $P\left(X_{i+q} < 2t_{k-1}+i \mid Y_{i,q-1}=d\right)$ can be obtained by looking at \cref{fig:proof_px_cond} and counting elements.
When $b_{t_{k-1}+i+q}$ is inserted, the elements on the main chain which are smaller than $a_{t_{k-1}+i}$ are $x_1$ to $x_{2t_{k-1}}$, $a_{t_{k-1}+1}$ to $a_{t_{k-1}+i-1}$ and $d$ elements out of $\{b_{t_{k-1}+i+1}, \dots, b_{t_{k-1}+i+q-1}\}$ which is a total of $2t_{k-1}+2i-1+d$ elements.
Combined with the fact that the main chain consists of $2t_{k-1} + 2i + 2q - 2$ elements smaller than $a_{t_{k-1}+i+q}$ we obtain the following formula
\begin{equation}
P\left(X_{i+q} < 2t_{k-1}+i \mid Y_{i,q-1}=d\right) = \frac{2t_{k-1}+2i+d}{2t_{k-1} + 2i + 2q - 1}
\label{eq:px_less}
\end{equation}

From that we can calculate
\begin{equation}
\begin{split}
&P(X_{i+q} \ge 2t_{k-1}+i \vert Y_{i,q-1}=d)\\
&=1-P(X_{i+q} < 2t_{k-1}+i \vert Y_{i,q-1}=d)\\
&=1-\frac{2t_{k-1}+2i+d}{2t_{k-1} + 2i + 2q - 1}\\
&=\frac{2t_{k-1} + 2i + 2q - 1-2t_{k-1}-2i-d}{2t_{k-1} + 2i + 2q - 1}\\
&=\frac{2q -d -1}{2t_{k-1} + 2i + 2q - 1}\\
\end{split}
\label{eq:px_ge}
\end{equation}

Now we have all the necessary ingredients to proof \cref{eq:Yiq_closed} using induction.

\begin{enumerate}

	\item Proof of \cref{eq:Yiq_closed} where $j=0$ using $\tilde{Y}_{i,q} = 0 \Leftrightarrow X_{i+q} \ge 2t_{k-1} + i \land \tilde{Y}_{i,q-1} = 0$
	\begin{equation}
	\resizebox{0.94\textwidth}{!}{$
		\begin{split}
		&P(X_{i+q} \ge 2t_{k-1} + i \land \tilde{Y}_{i,q-1} = 0)\\
		&=P\left(X_{i+q} \ge 2t_{k-1} + i \mid \tilde{Y}_{i,q-1} = 0\right) \cdot P(\tilde{Y}_{i,q-1} = 0)\\
		&\overset{\mathclap{Thm.\ref{th:Yi},(\ref{eq:px_ge})}}{=}\quad\frac{2q-1}{2t_{k-1}+2i+2q-1}\cdot\frac{(2q-2)!}{2^{q-1}0!(q-1)!}2^{q-1}\frac{(2t_{k-1}+2i-1)!}{(2t_{k-1}+2i+2q-3)!}\frac{(t_{k-1}+i+q-2)!}{(t_{k-1}+i-1)!}\\
		&=(2q-1)(2t_{k-1}+2i+2q-2)\cdot\frac{(2q-2)!}{2^{q-1}0!(q-1)!}2^{q-1}\frac{(2t_{k-1}+2i-1)!}{(2t_{k-1}+2i+2q-1)!}\frac{(t_{k-1}+i+q-2)!}{(t_{k-1}+i-1)!}\\
		&=(2q-1)2(t_{k-1}+i+q-1)\cdot\frac{(2q-2)!}{2^q0!(q-1)!}2^q\frac{(2t_{k-1}+2i-1)!}{(2t_{k-1}+2i+2q-1)!}\frac{(t_{k-1}+i+q-2)!}{(t_{k-1}+i-1)!}\\
		&=(2q-1)2\cdot\frac{(2q-2)!}{2^q0!(q-1)!}2^q\frac{(2t_{k-1}+2i-1)!}{(2t_{k-1}+2i+2q-1)!}\frac{(t_{k-1}+i+q-1)!}{(t_{k-1}+i-1)!}\\
		&=(2q-1)2\cdot\frac{q}{(2q)(2q-1)}\cdot\frac{(2q-0)!}{2^q0!(q-0)!}2^q\frac{(2t_{k-1}+2i-1)!}{(2t_{k-1}+2i+2q-1)!}\frac{(t_{k-1}+i+q-1)!}{(t_{k-1}+i-1)!}\\
		&=\frac{(2q-0)!}{2^q0!(q-0)!}2^q\frac{(2t_{k-1}+2i-1)!}{(2t_{k-1}+2i+2q-1)!}\frac{(t_{k-1}+i+q-1)!}{(t_{k-1}+i-1)!}\\
		&=P(\tilde{Y}_{i,q} = 0)
		\end{split}$}
	\end{equation}

	\item Proof of \cref{eq:Yiq_closed} where $j=q$ using $\tilde{Y}_{i,q} = q \Leftrightarrow X_{i+q} < 2t_{k-1}+i \land \tilde{Y}_{i,q-1}=q-1$
	\begin{equation}
	\resizebox{0.94\textwidth}{!}{$
		\begin{split}
		&P(X_{i+q} < 2t_{k-1}+i \land \tilde{Y}_{i,q-1}=q-1)\\
		&=P\left(X_{i+q} < 2t_{k-1}+i \mid \tilde{Y}_{i,q-1}=q-1\right)\cdot P(\tilde{Y}_{i,q-1}=q-1)\\
		&\overset{\mathclap{Thm.\ref{th:Yi},(\ref{eq:px_less})}}{=}\quad\frac{2t_{k-1}+2i+q-1}{2t_{k-1}+2i+2q-1}\cdot\frac{\left(q-1\right)!}{2^0\left(q-1\right)!0!}2^{q-1}\frac{\left(2t_{k-1}+2i+q-2\right)!}{\left(2t_{k-1}+2i+2q-3\right)!}\frac{\left(t_{k-1}+i+q-2\right)!}{\left(t_{k-1}+i-1\right)!}\\
		&=\left(2t_{k-1}+2i+q-1\right)\cdot\frac{\left(q-1\right)!}{2^0\left(q-1\right)!0!}2^{q}\frac{\left(2t_{k-1}+2i+q-2\right)!}{\left(2t_{k-1}+2i+2q-1\right)!}\frac{\left(t_{k-1}+i+q-1\right)!}{\left(t_{k-1}+i-1\right)!}\\
		&=\frac{\left(q-1\right)!}{2^0\left(q-1\right)!0!}2^{q}\frac{\left(2t_{k-1}+2i+q-1\right)!}{\left(2t_{k-1}+2i+2q-1\right)!}\frac{\left(t_{k-1}+i+q-1\right)!}{\left(t_{k-1}+i-1\right)!}\\
		&=\frac{\left(q\right)!}{2^0\left(q\right)!0!}2^{q}\frac{\left(2t_{k-1}+2i+q-1\right)!}{\left(2t_{k-1}+2i+2q-1\right)!}\frac{\left(t_{k-1}+i+q-1\right)!}{\left(t_{k-1}+i-1\right)!}\\
		&=P(\tilde{Y}_{i,q} = q)
		\end{split}$}
	\end{equation}

	\item Proof of \cref{eq:Yiq_closed} where $0<j<q$ using \\
	\resizebox{0.94\textwidth}{!}{$\tilde{Y}_{i,q} = j \Leftrightarrow (X_{i+q} < 2t_{k-1}+i \land \tilde{Y}_{i,q-1}=j-1)\lor (X_{i+q} \ge 2t_{k-1}+i \land \tilde{Y}_{i,q-1}=j)$}
	\begin{equation}
	\resizebox{0.94\textwidth}{!}{$
		\begin{split}
		&P(X_{i+q} < 2t_{k-1}+i \land \tilde{Y}_{i,q-1}=j-1)\\
		&\quad+P(X_{i+q} \ge 2t_{k-1}+i \land \tilde{Y}_{i,q-1}=j)\\
		&=P\left(X_{i+q} < 2t_{k-1}+i \mid \tilde{Y}_{i,q-1}=j-1\right)\cdot P(\tilde{Y}_{i,q-1}=j-1)\\
		&\quad+P\left(X_{i+q} \ge 2t_{k-1}+i \mid \tilde{Y}_{i,q-1}=j\right)\cdot P(\tilde{Y}_{i,q-1}=j)\\
		&\overset{\mathclap{Thm.\ref{th:Yi},(\ref{eq:px_less}),(\ref{eq:px_ge})}}{=}\quad\quad\frac{2t_{k-1}+2i+j-1}{2t_{k-1}+2i+2q-1}\cdot\frac{\left(2q-j-1\right)!}{2^{q-j}\left(j-1\right)!\left(q-j\right)!}2^{q-1}\frac{\left(2t_{k-1}+2i+j-2\right)!}{\left(2t_{k-1}+2i+2q-3\right)!}\frac{\left(t_{k-1}+i+q-2\right)!}{\left(t_{k-1}+i-1\right)!}\\
		&\quad+\frac{2q-j-1}{2t_{k-1}+2i+2q-1}\cdot\frac{\left(2q-j-2\right)!}{2^{q-j}\left(j-1\right)!\left(q-j\right)!}2^{q-1}\frac{\left(2t_{k-1}+2i+j-1\right)!}{\left(2t_{k-1}+2i+2q-3\right)!}\frac{\left(t_{k-1}+i+q-2\right)!}{\left(t_{k-1}+i-1\right)!}\\
		&=\left(2t_{k-1}+2i+j-1\right)\cdot\frac{\left(2q-j-1\right)!}{2^{q-j}\left(j-1\right)!\left(q-j\right)!}2^{q}\frac{\left(2t_{k-1}+2i+j-2\right)!}{\left(2t_{k-1}+2i+2q-1\right)!}\frac{\left(t_{k-1}+i+q-1\right)!}{\left(t_{k-1}+i-1\right)!}\\
		&\quad+\left(2q-j-1\right)\cdot\frac{\left(2q-j-2\right)!}{2^{q-j}\left(j-1\right)!\left(q-j\right)!}2^{q}\frac{\left(2t_{k-1}+2i+j-1\right)!}{\left(2t_{k-1}+2i+2q-1\right)!}\frac{\left(t_{k-1}+i+q-1\right)!}{\left(t_{k-1}+i-1\right)!}\\
		&=\left(\frac{\left(2q-j-1\right)!}{2^{q-j}\left(j-1\right)!\left(q-j\right)!}+\frac{\left(2q-j-1\right)!}{2^{q-j}\left(j-1\right)!\left(q-j\right)!}\right)2^{q}\frac{\left(2t_{k-1}+2i+j-1\right)!}{\left(2t_{k-1}+2i+2q-1\right)!}\frac{\left(t_{k-1}+i+q-1\right)!}{\left(t_{k-1}+i-1\right)!}\\
		&=\left(\frac{j}{2q-j}+\frac{2(q-j)}{2q-j}\right)\frac{\left(2q-j\right)!}{2^{q-j}j!\left(q-j\right)!}2^{q}\frac{\left(2t_{k-1}+2i+j-1\right)!}{\left(2t_{k-1}+2i+2q-1\right)!}\frac{\left(t_{k-1}+i+q-1\right)!}{\left(t_{k-1}+i-1\right)!}\\
		&=\frac{\left(2q-j\right)!}{2^{q-j}j!\left(q-j\right)!}2^{q}\frac{\left(2t_{k-1}+2i+j-1\right)!}{\left(2t_{k-1}+2i+2q-1\right)!}\frac{\left(t_{k-1}+i+q-1\right)!}{\left(t_{k-1}+i-1\right)!}\\
		&=P(\tilde{Y}_{i,q} = j)\\
		\end{split}$}
	\end{equation}

\end{enumerate}

From \cref{eq:Yiq_closed} we can derive \cref{th:Yi} using the \cref{eq:relYiYiq}.
\begin{equation}
\resizebox{\textwidth}{!}{$
\begin{split}
&P(Y_i = j)\\
&= P(\tilde{Y}_{i,t_k-t_{k-1}-i} + 2t_{k-1} + i - 1 = j)\\
&= P(\tilde{Y}_{i,t_k-t_{k-1}-i} = j - 2t_{k-1} - i + 1)\\
&= \frac{\left(2t_k-2t_{k-1}-2i-j+2t_{k-1}+i-1\right)!}{2^{t_k-t_{k-1}-i-j+2t_{k-1}+i-1}\left(j-2t_{k-1}-i+1\right)!\left(t_k-t_{k-1}-i-j+2t_{k-1}+1-1\right)!}\\
&\quad\cdot 2^{t_k-t_{k-1}-i}\frac{\left(2t_{k-1}+2i+j-2t_{k-1}-i+1-1\right)!}{\left(2t_{k-1}+2i+2t_k-2t_{k-1}-2i-1\right)!}
\frac{\left(t_{k-1}+i+t_k-t_{k-1}-i-1\right)!}{\left(t_{k-1}+i-1\right)!}\\
&=\frac{\left(2t_k-i-j-1\right)!}{2^{2^k-j-1}\left(-2t_{k-1}-1+j+1\right)!\left(2^k-j-1\right)!}\cdot 2^{t_k-t_{k-1}-i}\frac{\left(i+j\right)!}{\left(2t_k-1\right)!}\frac{\left(t_k-1\right)!}{\left(t_{k-1}+i-1\right)!}\\
&=2^{j-2t_{k-1}-i+1}\frac{\left(2t_k-i-j-1\right)!}{\left(-2t_{k-1}-1+j+1\right)!\left(2^k-j-1\right)!}\frac{\left(i+j\right)!}{\left(2t_k-1\right)!}\frac{\left(t_k-1\right)!}{\left(t_{k-1}+i-1\right)!}
\end{split}$}
\end{equation}

\subsection{Proof of \Cref{thm:upper_bound}}
\label{chap:appendix_upper_bound}

\begin{figure}
	\begin{center}
		\begin{tikzpicture}[scale=0.9]
		\filldraw (0.2, 0.8) circle (2pt) node [above] {\scalebox{0.7}{$x_1$}};
		\filldraw (1.4, 0.8) circle (2pt) node [above] {\scalebox{0.7}{$x_2$}};
		\filldraw (3.0, 0.8) circle (2pt) node [above] {\scalebox{0.7}{$x_{2t_{k-1}}$}};
		\filldraw (4.2, 0.8) circle (2pt) node [above] {\scalebox{0.7}{$a_{t_{k-1}+1}$}};
		\filldraw (5.4, 0.8) circle (2pt) node [above] {\scalebox{0.7}{$a_{t_{k-1}+2}$}};
		\filldraw (7.0, 0.8) circle (2pt) node [above] {\scalebox{0.7}{$a_{t_{k-1}+\frac{d}{2}}$}};
		\filldraw (8.2, 0.8) circle (2pt) node [above] {\scalebox{0.7}{$a_{t_{k-1}+\frac{d}{2}+1}$}};
		\filldraw (9.8, 0.8) circle (2pt) node [above] {\scalebox{0.7}{$a_{t_{k-1}+\left\lfloor\frac{3}{4}d\right\rfloor-1}\quad$}};
		\filldraw (11, 0.8) circle (2pt) node [above] {\scalebox{0.7}{$\quad a_{t_{k-1}+\left\lfloor\frac{3}{4}d\right\rfloor}$}};
		\filldraw (12.6, 0.8) circle (2pt) node [above] {\scalebox{0.7}{$a_{t_k}$}};
		\foreach \x in {0.2, 3.0, 4.2, 7.0, 9.8}
		\draw[-latex,shorten >=2pt] (\x, 0.8) -- (\x + 1.2, 0.8);
		\foreach \x in {1.4, 5.4, 8.2, 11}
		\draw[-latex,shorten >=2pt,dashed] (\x, 0.8) -- (\x + 1.6, 0.8);

		\filldraw (4.2 - 0.15, 0) circle (2pt) node [below] {\scalebox{0.7}{ $b_{t_{k-1}+1}$}};
		\filldraw (5.4 - 0.15, 0) circle (2pt) node [below] {\scalebox{0.7}{ $b_{t_{k-1}+2}$}};
		\filldraw (7.0 - 0.15, 0) circle (2pt) node [below] {\scalebox{0.7}{ $b_{t_{k-1}+\frac{d}{2}}$}};
		\filldraw (8.2 - 0.15, 0) circle (2pt) node [below] {\scalebox{0.7}{ $b_{t_{k-1}+\frac{d}{2}+1}$}};
		\filldraw (9.8 - 0.15, 0) circle (2pt) node [below] {\scalebox{0.7}{ $b_{t_{k-1}+\left\lfloor\frac{3}{4}d\right\rfloor-1}\quad$}};
		\filldraw (11 - 0.15, 0) circle (2pt) node [below] {\scalebox{0.7}{ $\quad b_{t_{k-1}+\left\lfloor\frac{3}{4}d\right\rfloor}$}};
		\filldraw (12.6 - 0.15, 0) circle (2pt) node [below] {\scalebox{0.7}{ $b_{t_k}$}};

		\foreach \x in {4.2,5.4,7.0,8.2,9.8,11,12.6}
		\draw[-latex,shorten >=2pt](\x - 0.15,0) -- (\x, 0.8);

		\draw[decorate,decoration={brace,mirror}] (8,-0.7) --node [below]{A} (10,-0.7);
		\draw[decorate,decoration={brace,mirror}] (10.8,-0.7) --node [below]{B} (12.8,-0.7);
		\end{tikzpicture}
	\end{center}
	\caption{Configuration where one batch is to be inserted.}
	\label{fig:conf_approx_Yi}
\end{figure}
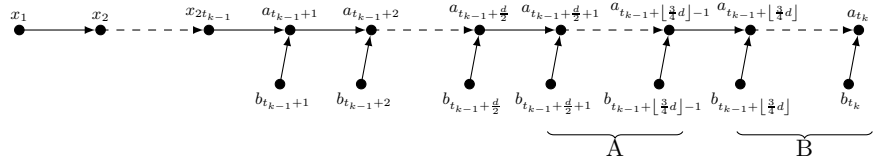

The exact probability that $b_{t_{k-1}+i}$ is inserted into $j$ elements is given by \cref{th:Yi}.
We are especially interested in the case of $b_{t_{k-1}+u}$ where $u = \lfloor \frac{t_k - t_{k-1}}{2} \rfloor$, 
because if we know $P(Y_u < m)$ then we can use that for all $q < u$ the probability of $b_{t_{k-1}+q}$ being inserted into less than $m$ elements is at least $P(Y_u < m)$, i.e. $P(Y_q < m) \ge P(Y_u < m)$.
This is because when $b_{t_{k-1}+i}$ is inserted into $m$ elements, then no matter which position it is inserted into, the next element, $b_{t_{k-1}+i-1}$, is inserted into at most $m$ elements.

However \cref{th:Yi} is hard to work with, so we approximate it with a binomial distribution.
For a given $k$ let $d=t_k-t_{k-1}$ be the number of elements that are inserted as part of the batch.
This configuration is illustrated in \cref{fig:conf_approx_Yi}.
Remember $u = \frac{t_k-t_{k-1}}{2} = \frac{d}{2}$.
To calculate into how many elements $b_{t_{k-1}+u} = b_{t_{k-1}+\frac{d}{2}}$ is inserted, we ask how many elements out of $b_{t_{k-1}+\left\lfloor\frac{3}{4}d\right\rfloor}$ to $b_{t_k}$ (marked as section B in \cref{fig:conf_approx_Yi}) are inserted between $a_{t_{k-1}+\frac{d}{2}+1}$ and $a_{t_{k-1}+\left\lfloor\frac{3}{4}d\right\rfloor-1}$ (marked as section A).

The rationale is that for each element from section B that is inserted into section A, $b_{t_{k-1}+u}$ is inserted into one less element.
As a lower bound for the probability that an element from section B is inserted into one of the positions in section A we use the probability that $b_{t_k}$ is inserted between $a_{t_k-1}$ and $a_{t_k}$ which is $\frac{1}{2t_k-1}$.

That is because if we assume that all $b_i$ with $i < t_k$ are inserted before inserting $b_{t_k}$, then $b_{t_k}$ is inserted into $2t_k-2$ elements, so the probability for each position is $\frac{1}{2t_k-1}$.
Since none of the $b_i$ with $i < t_k$ can be inserted between $a_{t_k-1}$ and $a_{t_k}$ because they are all smaller than $a_{t_k-1}$, the probability that $b_{t_k}$ is inserted between $a_{t_k-1}$ and $a_{t_k}$ does not change when we insert it first as the algorithm demands.

To calculate the probability that an element $b_{t_k-q}$ with $q>0$ is inserted into the rightmost position we assume that all $b_i$ with $i < t_k-q$ are inserted before inserting $b_{t_k-q}$.
Then $b_{t_k-q}$ is inserted into at most $2t_k-q-2$ elements, \ie the elements $x_1$ to $x_{2t_{k-1}}$, $a_{t_{k-1}+1}$ to $a_{t_k-q-1}$, $b_{t_{k-1}+1}$ to $b_{t_k-q-1}$ and at most $q$ elements out of $b_{t_k-q+1}$ to $b_{t_k}$.

Hence the probability for each position is greater than $\frac{1}{2t_k-q-1}$ which is greater than $\frac{1}{2t_k-1}$.
Since none of the $b_i$ with $i < t_k-q$ can be inserted to the right of $a_{t_k-q-1}$, the probability that $b_{t_k}-q$ is inserted into any of the positions between $a_{t_k-q-1}$ and $a_{t_k-q}$ remains unchanged when inserting the elements in the correct order.

The probability that an element is inserted at a specific position is monotonically decreasing with the index.
This is because if an element $b_i$ is inserted to the left of an element $a_{i-h}$ then $b_{i-h}$ is inserted into one more element than it would be if $b_i$ had been inserted to the right of $a_{i-h}$.
As a result any position further to the left is more likely than the right-most position, so we can use that as a lower bound.

There are $\left\lfloor\frac{d}{4}\right\rfloor - 1$ elements in section A, \ie there are at least $\left\lfloor\frac{d}{4}\right\rfloor$ positions where an element can be inserted.
Hence the probability that an element from section B is inserted into section A is at least $\frac{\left\lfloor\frac{d}{4}\right\rfloor}{2t_k-1}$ and consequently the probability that it is not inserted before $b_{t_{k-1}+u}$ is at least $\frac{\left\lfloor\frac{d}{4}\right\rfloor}{2t_k-1}$.
That is because all positions part of section A are after $a_{t_{k-1}+u}$.

Section B contains $\left\lceil\frac{d}{2}\right\rceil$ elements.
Using that and substituting $u = \frac{d}{2}$ we obtain the binomial distribution with the parameters $n_B=\left\lceil\frac{u}{2}\right\rceil$ and $p_B=\frac{\left\lfloor\frac{d}{4}\right\rfloor}{2t_k-1}$.
As a result we have 
\begin{equation}
p(j) = \binom{\left\lceil\frac{u}{2}\right\rceil}{q} (\frac{\lfloor \frac{u}{2} \rfloor}{2t_k-1})^q (\frac{2t_k-1 - \lfloor \frac{u}{2} \rfloor}{2t_k-1})^{\left\lceil\frac{u}{2}\right\rceil - q}
\end{equation}
with $q=2^k - 1 - j$, that by construction fulfills the property given in \Cref{eq:prop} for all $j_0$.
\begin{equation}
\sum_{j=0}^{j_0} p(j) \leq \sum_{j=0}^{j_0} P(Y_u = j) = P(Y_u \le j_0)
\label{eq:prop}
\end{equation}

\cref{fig:diff_Yi_approx} compares our approximation $p(j)$ with real distribution $P(Y_u = j)$.
We observe that the maximum of our approximation is further to the right than the one of the real distribution.

\ifplots
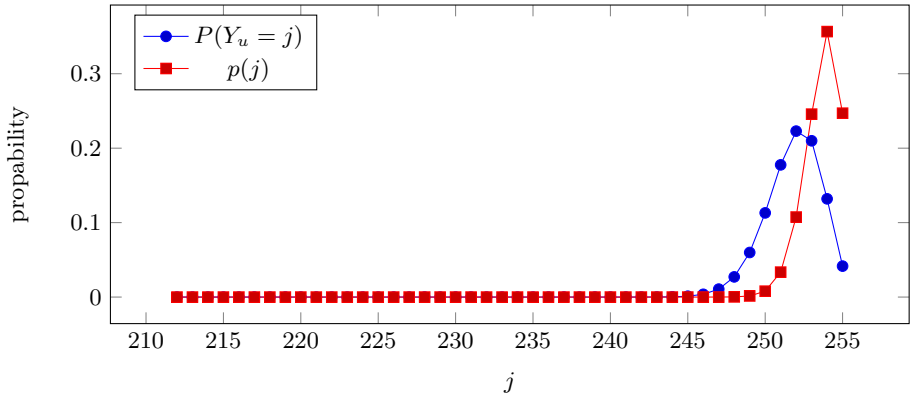
\begin{figure}[h]
	\begin{center}
		\begin{tikzpicture}
		\begin{axis}[xlabel=$j$,
		ylabel=propability,
		legend pos = north west,
		width=\textwidth,
		height=0.8*\axisdefaultheight]
		\addplot table [x=j, y=Y43, col sep=tab] {distr_Yapprox.csv};
		\addlegendentry{$P(Y_u = j)$}
		\addplot table [x=j, y=Y43Appr, col sep=tab] {distr_Yapprox.csv};
		\addlegendentry{$p(j)$}
		\end{axis}
		\end{tikzpicture}
	\end{center}
	\caption{Difference between the real distribution and our approximation for $k = 8$ and $u = 43$.}
	\label{fig:diff_Yi_approx}
\end{figure}
\fi

By using the approximation $P(Y_u = j) \approx p(j)$ we can calculate a lower bound for the median of $Y_\frac{t_k - t_{k-1}}{2}$
\begin{equation*}
\resizebox{\textwidth}{!}{$\displaystyle
\begin{split}
&2^k - 1 - \left\lfloor n_B \cdot p_B\right\rfloor\\
&= 2^k - 1 - \left\lfloor \left\lceil \frac{t_k - t_{k-1}}{4} \right\rceil \frac{\left\lfloor \frac{t_k - t_{k-1}}{4} \right\rfloor}{2t_k-1}\right\rfloor\\
& = 2^k - 1 - \left\lfloor \left( \frac{2^{k-2} + (-1)^k-3}{3} + \frac{1}{2} (-1)^k + \frac{1}{2} \right) \left( \frac{2^{k-2} + (-1)^k-3}{3} + \frac{1}{2} (-1)^k - \frac{1}{2} \right) \frac{1}{2t_k-1} \right\rfloor \\
& = 2^k - 1 - \left\lfloor \left( \frac{2^{k-2}}{3} + \frac{1}{6} (-1)^k + \frac{1}{2}\right) \left( \frac{2^{k-2}}{3} + \frac{1}{6} (-1)^k - \frac{1}{2}\right) \frac{1}{2t_k-1} \right\rfloor \\
& = 2^k - 1 - \left\lfloor  \left(\frac{2^{2k-4}}{9} + \frac{2^{k-2}}{9} (-1)^k + \frac{1}{36} - \frac{1}{4}\right)\frac{1}{2t_k-1} \right\rfloor \\
& = 2^k - 1 - \left\lfloor  \left(\frac{2^{2k-4}}{9} + \frac{2^{k-2}}{9} (-1)^k + \frac{1}{36} - \frac{1}{4}\right)\left(\frac{1}{2\frac{2^{k+1}+(-1)^k}{3}-1}\right) \right\rfloor \\
& = 2^k - 1 - \left\lfloor  \left(\frac{2^{2k-4}}{9} + \frac{2^{k-2}}{9} (-1)^k + \frac{1}{36} - \frac{1}{4}\right)\left(\frac{1}{2\frac{2^{k+1}}{3}}\pm\mathcal{O}(2^{-k})\right) \right\rfloor \\
& = 2^k - 1 - \left\lfloor  \left(\frac{2^{2k-4}}{9} + \frac{2^{k-2}}{9} (-1)^k + \frac{1}{36} - \frac{1}{4}\right)\frac{1}{2\frac{2^{k+1}}{3}}\pm\mathcal{O}(1) \right\rfloor \\
& = 2^k - 1 - \left\lfloor  \frac{2^{k-6}}{3} + \frac{1}{3} (-1)^k ]\pm\mathcal{O}(1) \right\rfloor \\
& \in 2^k - 1 - \frac{2^{k-6}}{3}+\mathcal{O}(1)
\end{split}$}
\end{equation*}

This tells us that with a probability $\geq 50\% $, $b_{t_{k-1}+u}$ is inserted into $2^k - 1 - \frac{2^{k-6}}{3} \pm\mathcal{O}(1)$ or less elements.
In conclusion all $b_i$ with $i \le u = \frac{t_k - t_{k-1}}{2}$ are inserted into less than $2^k - 1 - \frac{2^{k-6}}{3} \pm\mathcal{O}(1)$ elements with a probability $\geq 50\% $.

Using that result we can calculate a better upper bound for the average case performance of the entire algorithm.

According to Knuth \cite{Knuth} in its worst case MergeInsertion requires $W(n) = n \log n - (3 - \log 3)n + n(y+1-2^y) + \mathcal{O}(\log n)$ comparisons where $y = y(n) = \left\lceil \log(3n/4) \right\rceil - \log(3n/4) \in \lbrack 0, 1)$. 

We calculate the number of comparisons required in the average case in a similar fashion to \cite{EdelkampWW18}. Recall \cref{eq:Fn} which is the number of comparisons required by the algorithm.
\begin{equation*}
F(n) = \left\lfloor \frac{n}{2} \right\rfloor + F\left(\left\lfloor \frac{n}{2} \right\rfloor\right) + G\left(\left\lceil \frac{n}{2} \right\rceil\right)
\end{equation*}
$G(m)$ corresponds to the work done in the third step of the algorithm and is given by
\begin{equation*}
G(m) = (k_m - \alpha_m)(m - t_{k_m-1}) + \sum_{1 \le k < k_m} (k - \beta_k) \left(t_k-t_{k-1}\right)
\label{eq:micost_G}
\end{equation*}
where $t_{k_m-1} \le m < t_{k_m}$ and $\alpha_m, \beta \in \lbrack 0, 1\rbrack$.
Inserting an element $b_i$ with $t_{k_{i-1}} < i \le t_{k_i}$ requires at most $k_i$ comparisons.
However, since we are looking at the average case we need to consider that in some cases $b_i$ can be inserted using just $k_i-1$ comparisons.
This is reflected by $\alpha_m $ and $\beta_k$, the first of which has already been studied by \cite{EdelkampWW18}.

To estimate the cost of an insertion we use the formula $T_{\text{InsAvg}}(m) = \lceil \log m \rceil + 1 - \frac{2^{\lceil \log m \rceil}}{m}$ by \cite{EdelkampWW18}.
Technically this formula is only correct if the probability of an element being inserted is the same for each position.
This is not the case with MergeInsertion.
Instead the probability is monotonically decreasing with the index.
Binary insertion can be implemented to take advantage of this property, as explained in \cref{chap:BinIns}, in which case $T_{\text{InsAvg}}(m)$ acts as an upper bound on the cost of an insertion.

Using our result from above that on average $\frac{1}{4}$ of the elements are inserted in less than $2^k - 1 - \frac{2^{k-4}}{9} \pm\mathcal{O}(1)$ elements we can calculate $\beta_k$ as the difference of the cost of an insertion in the worst-case ($k$) and in the average case.

\begin{equation*}
\begin{split}
\beta_k
&\ge k - \left(\frac{3}{4}T_{\text{InsAvg}}\left(2^k\right) + \frac{1}{4}T_{\text{InsAvg}}\left(2^k - \frac{2^{k-6}}{3} \pm \mathcal{O}(1)\right)\right)\\
&= k - \left(\frac{3}{4}\left(k + 1 - \frac{2^k}{2^k}\right) + \frac{1}{4}\left(k + 1 - \frac{2^k}{2^k - \frac{2^{k-6}}{3} \pm \mathcal{O}(1)}\right)\right)\\
&= -1 +\frac{3}{4} + \frac{1}{4}\cdot\frac{1}{1 - \frac{1}{1 - \frac{2^{-6}}{3}} \pm \mathcal{O}(2^{-k})}\\
&= -\frac{1}{4} + \frac{1}{4}\cdot\frac{1}{1-\frac{1}{192}} \pm \mathcal{O}(2^{-k})\\
&= -\frac{1}{4} + \frac{1}{4}\cdot\frac{1}{\frac{191}{192}} \pm \mathcal{O}(2^{-k})\\
&= -\frac{1}{4} + \frac{1}{4}\cdot\frac{192}{191} \pm \mathcal{O}(2^{-k})\\
&= \frac{1}{764} \pm \mathcal{O}(2^{-k})\\
\end{split}
\end{equation*}

Combining this with \cref{eq:micost_G} we can calculate the difference between the worst-case and the average-case as
\begin{align}\allowdisplaybreaks
\nonumber &G_\text{worst-case}(m) - G_\text{average-case}(m)\\
\nonumber& = \begin{aligned}[t]& k_m(m - t_{k_m-1}) + \sum_{1 \le k < k_m} k \left(t_k-t_{k-1}\right) \\
\nonumber    &- (k_m - \alpha_m)(m - t_{k_m-1}) - \sum_{1 \le k < k_m} (k - \beta_k) \left(t_k-t_{k-1}\right)\end{aligned}\\
\nonumber&= \alpha_m(m - t_{k_m-1}) + \sum_{1 \le k < k_m} \beta_k\left(t_k-t_{k-1}\right)\\
\nonumber&\ge \alpha_m(m - t_{k_{m-1}}) + \sum_{1 \le k < k_m} (\frac{1}{764} \pm \mathcal{O}(2^{-k})) \left(t_k-t_{k-1}\right)\\
\nonumber&= \alpha_m(m - t_{k_m-1}) + \frac{1}{764}(t_{k_{m-1}} - t_1) \pm \mathcal{O}(\log m)\\
\nonumber&= \alpha_m(m - t_{k_m-1}) + \frac{1}{764}t_{k_{m-1}} \pm \mathcal{O}(\log m)\\
\nonumber&= \alpha_m(m - t_{k_m-1}) + \frac{1}{764}\frac{2^k_m + (-1)^{k_m-1}}{3} \pm \mathcal{O}(\log m)\\
&= \alpha_m(m - t_{k_m-1}) + \frac{1}{764}\frac{2^{k_m}}{3} \pm \mathcal{O}(\log m)
\label{eq:gwc-gac}
\end{align}

By writing $m$ as $m = 2^{l_m - \log 3 + x}$ with $x \in \lbrack 0, 1)$ we get $l_m = \left\lfloor \log 3m \right\rfloor$.
To approximate $k_m$ with $l_m$ we need to show that $k_m \ge l_m$.
Recall that $t_{k_m-1} \le m < t_{k_m}$.
For all $t_{k_m-1} < m < t_{k_m}$ we have
\begin{equation*}
\frac{2^{k_m} + (-1)^{k_m-1}}{3} < m < \frac{2^{k_m+1} + (-1)^{k_m}}{3}
\end{equation*}
Since $m \in \mathbb{N}$ and $t_k \in \mathbb{N}$ adding/subtracting $\frac{1}{3}$ does not alter the relation, so we obtain
\begin{equation*}
\frac{2^{k_m}}{3} < m < \frac{2^{k_m+1}}{3}
\end{equation*}
which resolves to
\begin{equation*}
k_m < \log 3n < k_m + 1
\end{equation*}
Thus $k_m = \left\lfloor \log 3m \right\rfloor = l_m$.

For $m = t_{k_m-1}$ we get
\begin{equation*}
\begin{aligned}
&& \frac{2^{k_m}+(-1)^{k_m-1}}{3} &= m\\
\Longleftrightarrow && 2^{k_m} &= 3m + (-1)^{k_m}\\
\Longleftrightarrow && k_m &= \log\left(3m + (-1)^{k_m}\right)
\end{aligned}
\end{equation*}

If $k_m = \log\left(3m + 1\right)$ that resolves to $k_m = \log\left(3m + 1\right) > \log\left(3m\right) > \left\lfloor \log 3m \right\rfloor = l_m$.

If instead $k_m = \log\left(3m - 1\right)$ using $k_m \in \mathbb{N}$ we have $k_m = \left\lfloor \log (3m - 1) \right\rfloor $ and for all $m\ge 1$ this is equal to $\left\lfloor \log 3m \right\rfloor = l_m$.

Hence in all cases $l_m \le k_m$ holds.
Therefore we can replace $k_m$ with $l_m$ in \cref{eq:gwc-gac}:
\begin{equation*}
G_\text{worst-case}(m) - G_\text{average-case}(m) \ge \alpha_m(m - t_{k_m-1}) + \frac{1}{764}\frac{2^{l_m}}{3} \pm \mathcal{O}(\log m)
\end{equation*}

From \cite{EdelkampWW18} we know that the $\alpha_m(m - t_{k_m-1})$ term can be approximated with $\left(m - 2^{l_m - \log 3}\right)\left(\frac{2^{l_m}}{m + 2^{l_m - \log 3}}-1\right)$.
\begin{align*}
&G_\text{worst-case}(m) - G_\text{average-case}(m) \\
&\ge \left(m - 2^{l_m - \log 3}\right)\left(\frac{2^{l_m}}{m + 2^{l_m - \log 3}}-1\right) + \frac{1}{764}\frac{2^{l_m}}{3} \pm \mathcal{O}(\log m)
\end{align*}

Now we calculate
\begin{equation}
\begin{split}
S(n) &= F_\text{worst-case}(m) - F_\text{average-case}(m)\\
&= \left\lfloor \frac{n}{2} \right\rfloor + F_\text{worst-case}\left(\left\lfloor \frac{n}{2} \right\rfloor\right) + G_\text{worst-case}\left(\left\lceil \frac{n}{2} \right\rceil\right)\\
&\quad - \left\lfloor \frac{n}{2} \right\rfloor - F_\text{average-case}\left(\left\lfloor \frac{n}{2} \right\rfloor\right) - G_\text{average-case}\left(\left\lceil \frac{n}{2} \right\rceil\right)\\
&= S(\left\lfloor \frac{n}{2} \right\rfloor) + G_\text{worst-case}\left(\left\lceil \frac{n}{2} \right\rceil\right) - G_\text{average-case}\left(\left\lceil \frac{n}{2} \right\rceil\right)\\
&\ge S(\left\lfloor \frac{n}{2} \right\rfloor) + \left(m - 2^{l_m - \log 3}\right)\left(\frac{2^{l_m}}{m + 2^{l_m - \log 3}}-1\right) + \frac{1}{764}\frac{2^{l_m}}{3} \pm \mathcal{O}(\log m)\\
\end{split}
\label{eq:avg_savings}
\end{equation}

We split $S(n)$ into $S_\alpha(n) + S_\beta(n)$ with
\begin{equation*}
\begin{array}{rl}
S_\alpha(n) &\ge S_\alpha(\left\lfloor \frac{n}{2} \right\rfloor) + \left(m - 2^{l_m - \log 3}\right)\left(\frac{2^{l_m}}{m + 2^{l_m - \log 3}}-1\right)\\
S_\beta(n) &\ge S_\beta(\left\lfloor \frac{n}{2} \right\rfloor) + \frac{1}{764}\frac{2^{l_m}}{3} \pm \mathcal{O}(\log m)
\end{array}
\end{equation*}

From \cite{EdelkampWW18} we know $S_\alpha(n) \ge \left(n - 2^{l_n - \log 3}\right)\left(\frac{2^{l_n}}{n + 2^{l_n - \log 3}}-1\right) + \mathcal{O}(1)$.

For $S_\beta(n)$ we obtain
\begin{equation*}
\begin{split}
S_\beta(n) &\ge \sum_{i= 1}^{l_n - 1}\frac{2^i}{764\cdot 3}\pm\mathcal{O}(\log 2^i)\\
&= \frac{2^{l_n}}{2292} \pm \mathcal{O}(\log^2 n)
\end{split}
\end{equation*}

We can represent $n$ as $2^{k - \log 3 + x_n}$ with $x_n \in \lbrack 0, 1)$. This leads to
\begin{equation*}
\resizebox{\textwidth}{!}{$
\begin{split}
\frac{S(n)}{n} &= \frac{S_\alpha(n) + S_\beta(n)}{n}\\
&= \frac{2^{k - \log 3 + x_n} - 2^{k - \log 3}}{2^{k - \log 3 + x_n}}\left(\frac{2^k}{2^{k - \log 3 + x_n} + 2^{k - \log 3}}-1\right) + \frac{2^k}{2292\cdot 2^{k - \log 3 + x_n}} \pm \mathcal{O}(\frac{\log^2 n}{n})\\
&= (1-2^{-x_n})\left(\frac{3}{2^{x_n}+1}-1\right)+\frac{2^{\log 3 - x_n}}{2292} \pm \mathcal{O}(\frac{\log^2 n}{n})
\end{split}$}
\end{equation*}

By writing $F(n)$ as $F(n) = n \log n - c(x_n)\cdot n \pm \mathcal{O}(\log^2 n)$ we get
\begin{equation*}
\begin{split}
c(x_n) &\ge -\frac{(F(n) - n\log n)}{n}\\
&= -\frac{(W(n) - S(n) - n\log n)}{n}\\
&= (3 - \log 3) - (y + 1 - 2^y) + (1-2^{-x_n})\left(\frac{3}{2^{x_n}+1}-1\right)+\frac{2^{\log 3 - x_n}}{2292}\\
\end{split}
\end{equation*}
With $y = 1 - x_n$ we obtain \Cref{thm:upper_bound}.

\section{Details on Computing the Exact Number of Comparisons}
\label{chap:appendix_exact}

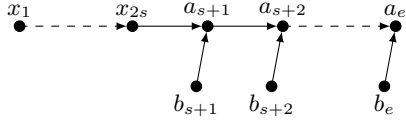
\begin{figure}[ht]
	\centering
	\begin{tikzpicture}
	\filldraw (0, 0.8) circle (2pt) node [above] {$x_1$}
	(1.5, 0.8) circle (2pt) node [above] {$x_{2s}$}
	(2.5, 0.8) circle (2pt) node [above] {$a_{s+1}$}
	(3.5, 0.8) circle (2pt) node [above] {$a_{s+2}$}
	(5, 0.8) circle (2pt) node [above] {$a_{e}$};
	
	\draw[-latex,shorten >=2pt](1.5, 0.8) -- (2.5, 0.8);
	\draw[-latex,shorten >=2pt](2.5, 0.8) -- (3.5, 0.8);
	\draw[-latex,shorten >=2pt,dashed] (0, 0.8) -- (1.5, 0.8);
	\draw[-latex,shorten >=2pt,dashed] (3.5, 0.8) -- (5, 0.8);
	
	\filldraw (2.5 - 0.15, 0) circle (2pt) node [below] {$b_{s+1}$}
	(3.5 - 0.15, 0) circle (2pt) node [below] {$b_{s+2}$}
	(5 - 0.15, 0) circle (2pt) node [below] {$b_{e}$};
	
	\draw[-latex,shorten >=2pt](2.5 - 0.15,0) -- (2.5, 0.8);
	\draw[-latex,shorten >=2pt](3.5 - 0.15,0) -- (3.5, 0.8);
	\draw[-latex,shorten >=2pt](5 - 0.15,0) -- (5, 0.8);
	\end{tikzpicture}
	\caption{Configuration where one batch of $e-s$ elements, $b_{s+1}$ to $b_e$, remains to be inserted.}
	\label{fig:conf_batch_comp}
\end{figure}

The code for calculating $F(n)$ and $G(n)$ is shown in \Cref{alg:computef} and \Cref{alg:computeg} respectively.

$\texttt{Cost}(s, e)$ is the number of comparisons required for inserting the batch of elements that consists of $b_{s+1}$ to $b_e$.
Such a configuration can be seen in \cref{fig:conf_batch_comp}.
$\texttt{Cost}(s, e)$ is computed by calculating the external path length of the decision tree and dividing by the number of leaves.
To improve performance we apply the following optimization:
We collapse ``identical'' branches of the decision tree.
E.g. whether $b_e$ is inserted between $x_1$ and $x_2$ or between $x_2$ and $x_3$ does not influence the number of comparisons required to insert the subsequent elements. So we can neglect that difference.
However, if $b_e$ is inserted between $a_{e-1}$ and $a_e$ then the next element (and all thereafter) is inserted into one less element.
So this is a difference we need to acknowledge.
Same if an element is inserted between any $a_i$ and $a_{i+1}$.
By the time we insert $b_i$ the element inserted between $a_i$ and $a_{i+1}$ is known to be larger than $b_i$ and thus is no longer part of the main chain, resulting in $b_i$ being inserted into one element less.
In conclusion that means that our algorithm needs to keep track of the elements inserted between any $a_i$ and $a_{i+1}$ as well as those inserted at any position before $a_{s+1}$ as two branches of the decision tree that differ in any of these cannot be collapsed.
\Cref{alg:cost} shows how this is implemented.

\begin{algorithm}
	\begin{algorithmic}[1]
		\Procedure{ComputeF}{$n$}
		\If{$n=1$}
		\State \Return $0$
		\Else
		\State \Return $\left\lfloor \frac{n}{2} \right\rfloor + $\Call{ComputeF}{$\left\lfloor \frac{n}{2} \right\rfloor$}$ + $\Call{ComputeG}{$\left\lceil \frac{n}{2} \right\rceil$}
		\EndIf
		\EndProcedure
	\end{algorithmic}
	\caption{Computation of $F(n)$}
	\label{alg:computef}
\end{algorithm}

\begin{algorithm}
	\begin{algorithmic}[1]
		\Procedure{ComputeG}{$n$}
		\State $k \gets 2$
		\State $c \gets 0$
		\While {$t_k < n$}
		\State $c \gets c + $\Call{Cost}{$t_{k-1}$, $t_k$}
		\State $k \gets k+1$
		\EndWhile
		\State $c \gets c + $\Call{Cost}{$t_{k-1}$, $n$}
		\State \Return $c$
		\EndProcedure
	\end{algorithmic}
	\caption{Computation of $G(n)$}
	\label{alg:computeg}
\end{algorithm}

\algblockdefx[TIMES]{Times}{EndTimes}[1]{\textbf{repeat} #1 \textbf{times}}{\textbf{end}}
\begin{algorithm}
	\begin{algorithmic}[1]
		\Procedure{Cost}{$s$, $e$}
		\State $r \gets e-s$ \Comment{next element to be inserted is $b_{r}$}
		\State $q_1 \gets 2s$ \Comment{number of elements on the main chain that are $<a_{s+1}$}
		\State $q_2, \dots, q_r \gets 0$ \Comment{$q_i$ is the number of elements between $a{s+i-1}$ and $a_{s+i}$}
		\State $(p, l) \gets $\Call{CostInsert}{$r$, $q_1$, ..., $q_r$}
		\State \Return $\frac{p}{l}$
		\EndProcedure
		\State
		\Procedure{CostInsert}{$r$, $q_1$, ..., $q_r$}
		\If{$r=0$}
		\State\Return $(0, 1)$ \Comment{We reached a leave}
		\EndIf
		\State $elements \gets r - 1 + \sum q_i$ \Comment{number of elements $b_{r}$ is inserted into}
		\State $k \gets \lceil \log (elements + 1) \rceil$
		\State $cheap\_insertions \gets 2^k - elements - 1$
		\State $p \gets 0$ \Comment{external path length}
		\State $l \gets 0$ \Comment{number of leaves}

		\State $index \gets 0$\Comment{We iterate over all indices where $b_{r}$ can be inserted}
		\ForAll{$0<i\le r$}
		\State $(p_c, l_c) \gets $\Call{CostInsert}{$r-1$, $q_1$, ..., $q_{i-1}$, $q_i+1$, $q_{i+1}$, ..., $q_{r-1}$}
		\Times{$q_i+1$} \Comment{$q_i+1$ positions between $a_{s+i-1}$ and $a_{s+i}$}
		\If{$index < cheap\_insertions$}
		\State $p \gets p + p_c + (k-1)\cdot l_c$
		\Else
		\State $p \gets p + p_c + k\cdot l_c$
		\EndIf
		\State $l \gets l + l_c$
		\State $index \gets index + 1$
		\EndTimes
		\EndFor
		\State\Return $(p, l)$
		\EndProcedure
	\end{algorithmic}
	\caption{Computation of $\texttt{Cost}(s, e)$}
	\label{alg:cost}
\end{algorithm}
\clearpage
\section{Implementing MergeInsertion}
\label{chap:impl}

To perform experiments we first need to implement the algorithm.
For the purpose of our implementation we assume that each element is unique.
This condition is easy to fulfill for synthetic test data.
You can see our implementation in \Cref{alg:mi}.
We now go over some of the key challenges when implementing MergeInsertion.

\begin{algorithm}
	\begin{algorithmic}[1]
		\Procedure{MergeInsertion}{$d: \text{array of } n \text{ elements}$}
		\State \text{Step 1: Pairwise comparison}
		\ForAll{$1 \le i \le \left\lfloor\frac{n}{2}\right\rfloor$} \Comment{Split into larger and smaller half}
		\State $a_i \gets \max \left\lbrace d_i, a_{i+\left\lfloor\frac{n}{2}\right\rfloor}\right\rbrace$
		\State $b_i \gets \min \left\lbrace d_i, a_{i+\left\lfloor\frac{n}{2}\right\rfloor}\right\rbrace$
		\EndFor
		\If{$n \bmod 2 = 1$}
		\State $b_{\left\lceil\frac{n}{2}\right\rceil} \gets d_n$
		\EndIf

		\State \text{Step 2: Recursion and Renaming}
		\State $m \gets \left\lbrace \left(a_i, b_i \right) \mid 1 \le i \le \left\lfloor\frac{n}{2}\right\rfloor \right\rbrace$\label{alg:mi:makemap} \Comment{Store mapping}
		\State $a \gets $\Call{MergeInsertion}{$a$}
		\ForAll{$1 \le i \le \left\lfloor\frac{n}{2}\right\rfloor$} \label{alg:mi:permute}\Comment{Permute smaller half}
		\State $b_i \gets e \text{ where } \left(a_i, e \right) \in m$
		\EndFor

		\State \text{Step 3: Insertion}
		\State $d \gets b_1, a_1, \dots, a_{\left\lfloor\frac{n}{2}\right\rfloor}$
		\State $k \gets 2$
		\While{$t_{k-1} < \left\lceil\frac{n}{2}\right\rceil$}
		\State $m \gets \min \left\lbrace t_k, \left\lceil\frac{n}{2}\right\rceil\right\rbrace$ \Comment{first element of the batch}
		\State $u \gets t_{k-1} + m$\label{alg:mi:makeu}\Comment{position of $a_m$ in $d$}
		\For{$i$ in $m$ down to $t_{k-1}+1$}
		\State $d \gets $\Call{BinaryInsertion}{$b_i$, $d_1$, ..., $d_{u-1}$}, $d_u$, ..., $d_{2m+t_{k-1} - i}$
		\While{$d_u \ne a_{i-1}$}\Comment{adjust $u$}
		\State $u \gets u - 1$\label{alg:mi:updateu}
		\EndWhile
		\EndFor
		\State $k \gets k + 1$
		\EndWhile
		\State\Return $d$
		\EndProcedure
	\end{algorithmic}
	\caption{MergeInsertion}
	\label{alg:mi}
\end{algorithm}

\begin{enumerate}
	\item MergeInsertion requires elements to be inserted into arbitrary positions.
	When using a simple array to store the elements this operation requires moving $\mathcal{O}(n)$ elements.
	Since MergeInsertion inserts each element exactly once this results in a complexity of $\mathcal{O}(n^2)$.
	To avoid this we store the elements in a custom data structure inspired by the Rope data structure\cite{Rope} used in text processing.
	Being based on a tree it offers $\mathcal{O}(\log n)$ performance for lookup, insertion and deletion operations, thus putting our Algorithm in $\mathcal{O}(n\log^2 n)$.

	\item In the second step of the algorithm we need to rename the $b_i$ after the recursive call.
	Our chosen solution is to store which $a_i$ corresponds to which $b_i$ in a hash map(line \ref{alg:mi:makemap}) before the recursive call and use the information to reorder the $b_i$ afterwards(line \ref{alg:mi:permute}).
	The disadvantage of this solution is that it requires each element to be unique and the hash map might introduce additional comparisons.

	An alternative would be to have the recursive call generate the permutation it applies to the larger elements and then apply that to the smaller ones.
	That is a cleaner solution as it does not require the elements to be unique and it avoids potentially introducing additional comparisons.
	It is also potentially faster, though not by much.
	However, we stuck with using a hash map as that solution is easier to implement.

	\item In the insertion step we need to know into how many elements a specific $b_i$ is inserted.
	For $b_{t_k}$ this is $2^k-1$ elements.
	However, for other elements that number can be smaller depending on where the previous elements have been inserted.
	To account for that we create the variable $u$ in line \ref{alg:mi:makeu}.
	It holds the position of the $a_i$ corresponding to the element $b_i$ that is inserted next.
	Thus $b_i$ is inserted into $u-1$ elements (since $b_i < a_i$).
	After the insertion of $b_i$, we decrease $u$ in line \ref{alg:mi:updateu} until it matches the position of $a_{i-1}$, which is what we want as $b_{i-1}$ is the next element to be inserted.
	This step also makes use of the requirement that each element is unique.

	At this point we have to be aware that testing whether the element at position $u$ is $a_{i-1}$ might introduce additional comparisons to the algorithm.
	This is acceptable because we do not count these comparisons.
	Also these are not necessary.
	We could keep track of the positions of the elements $a_i$ however we choose not to, in order to keep the implementation simple.

\end{enumerate}

\end{document}